\documentclass{iopjournal}

\usepackage[T1]{fontenc}
\usepackage{url}
\usepackage{amsmath,amsfonts,amsthm}
\usepackage[capitalise]{cleveref}       
\usepackage{graphicx}
\usepackage[dvipsnames]{xcolor}
\usepackage{tikz}
\usetikzlibrary{arrows.meta,positioning,calc,intersections,shapes,angles,patterns,quotes,backgrounds}
\usepackage{circuitikz}
\ctikzset{tripoles/pmos style/emptycircle}
\ctikzset{tripoles/mos style/arrows}
\usepackage{siunitx}
\usepackage{tikzscale}
\usepackage[acronym,nohypertypes={acronym}]{glossaries}
\usepackage{orcidlink}
\usepackage{pgfplots}
\usepackage{tabularray}
\DeclareUnicodeCharacter{2212}{−}
\usepgfplotslibrary{groupplots,dateplot}
\pgfplotsset{compat=newest}
\usepackage{cite}
\usepackage{float}
\usepackage{enumitem}
\usepackage[defaultcolor=myred]{changes}

\graphicspath{Figures}
\DeclareGraphicsExtensions{.pdf,.jpeg,.png,.tikz}

\definecolor{myblue}{RGB}{0,114,178}
\definecolor{myred}{RGB}{213,94,0}
\definecolor{mygreen}{RGB}{0,158,115}
\definecolor{myorange}{RGB}{230,159,0}
\definecolor{mycyan}{RGB}{86,180,233}
\definecolor{myyellow}{RGB}{240,228,66}
\definecolor{mypink}{RGB}{204,121,167}
\definecolor{mygray}{RGB}{204,204,204}

\tikzset{
  sum/.style={draw, circle, inner sep=1pt, minimum size=10pt, thick},
  signal/.style={-{Latex[length=3mm, width=2mm]}, thick},
  modsignal/.style={-{Latex[open, length=3mm, width=2mm]}, thick, dashed},
  fblock/.style={draw, thick, minimum height=35, minimum width=40, align=center, rounded corners=5pt},
  sig pic/.pic={
    \draw[thick, black] (-0.5,-0.5) .. controls (0.3,-0.5) and (-0.3,0.5) .. (0.5,0.5);
  },
  sblock/.style={draw, thick, minimum height=35, minimum width=40, align=center, rounded corners=5pt,
  path picture={
  \pic at (path picture bounding box.center) {sig pic};
  }
  }
}

\theoremstyle{definition}

\DeclareSIUnit{\sample}{S}

\usepackage[precision=2, unit=mm]{lengthconvert}

\renewcommand{\iff}{\Leftrightarrow}

\newacronym{dpi}{DPI}{differential-pair integrator}
\newacronym{dic}{DIC}{dynamic input conductances}
\newacronym{iandf}{I\&F}{integrate-and-fire}
\newacronym{cpg}{CPG}{central pattern generator}
\newacronym{pcb}{PCB}{printed circuit board}
\newacronym{dac}{DAC}{digital-to-analog converter}
\newacronym{cmos}{CMOS}{Complementary Metal-Oxide-Semiconductor}
\newacronym{fet}{FET}{field-effect transistor}
\newacronym{aer}{AER}{Address Event Representation}
\newacronym{adp}{ADP}{after-depolarization potential}
\newacronym{cv}{CV}{coefficient of variation}
\newacronym{iqr}{IQR}{interquartile range}
\usepackage{lmodern}

\newlength{\singlefigwidth}
\newlength{\doublefigwidth}
\begin{document}

\articletype{Original version submitted on November 30, 2025. Revised on April 8, 2026.}

\title{A Neuromodulable Current-Mode Silicon Neuron for Robust and Adaptive Neuromorphic Systems}

\author{Loris~Mendolia$^{1,*}$\orcid{0000-0003-0270-5736},
Chenxi~Wen$^2$\orcid{0009-0008-5204-5709},
Elisabetta~Chicca$^3$\orcid{0000-0002-5518-8990},
Giacomo~Indiveri$^2$\orcid{0000-0002-7109-1689},
Rodolphe~Sepulchre$^4$\orcid{0000-0002-7047-3124},
Jean-Michel~Redouté$^1$\orcid{0000-0001-9612-4312}
and Alessio Franci$^{1,5,*}$\orcid{0000-0002-3911-625X}}

\affil{$^1$Department of Electrical Engineering and Computer Science, University of Liège, Liège, Belgium}

\affil{$^2$Institute of Neuroinformatics, University of Zurich \& ETH Zurich, Zurich, Switzerland}

\affil{$^3$Bio-Inspired Circuits and Systems Lab, Zernike Institute for Advanced Materials \& Groningen Cognitive Systems and Materials Center, University of Groningen, Groningen, The Netherlands}

\affil{$^4$Department of Engineering, University of Cambridge, Cambridge, U.K. \& Department of Electrical Engineering, KU Leuven, Leuven, Belgium}

\affil{$^5$WEL Research Institute, Wavre, Belgium}

\affil{$^*$Authors to whom any correspondence should be addressed.}

\email{\href{mailto:lmendolia@uliege.be}{lmendolia@uliege.be}, \href{mailto:afranci@uliege.be}{afranci@uliege.be}}


\begin{abstract}
Neuromorphic engineering makes use of mixed-signal analog and digital circuits to directly emulate the computational principles of biological brains.
Such electronic systems offer a high degree of adaptability, robustness, and energy efficiency across a wide range of tasks, from edge computing to robotics.
Within this context, we investigate a key feature of biological neurons: their ability to carry out robust and reliable computation by adapting their input responses and spiking patterns to context through neuromodulation.
Achieving analogous levels of robustness and adaptation in neuromorphic circuits through modulatory mechanisms is a largely unexplored path.
We present a novel current-mode neuron design that supports robust neuromodulation with minimal model complexity, compatible with standard CMOS technologies.
We first introduce a mathematical model of the circuit and provide tools to analyze and tune the neuron behavior; we then demonstrate both theoretically and experimentally the biologically plausible neuromodulation adaptation capabilities of the circuit over a wide range of parameters.
All theoretical predictions were verified in experiments on a low-power \SI{180}{\nano\meter} CMOS implementation of the proposed neuron circuit.
Due to the analog underlying feedback structure, the proposed adaptive neuromodulable neuron exhibits a high degree of robustness, flexibility, and scalability across operating ranges of currents and temperatures, making it a perfect candidate for real-world neuromorphic applications.
\end{abstract}

\section{Introduction}
The remarkable efficiency and adaptability of the human brain emerge not only from the dynamics of spiking neurons and plastic synapses, but also from a cellular regulation mechanism known as neuromodulation~\cite{bargmann_beyond_2012,marder_neuromodulation_2012,avery_neuromodulatory_2017,drion2019celullar}. Neuromodulators, such as dopamine and serotonin, shape neuronal behavior over timescales ranging from seconds to minutes by adjusting excitability, firing thresholds and spiking patterns, and by regulating rhythmic activity~\cite{mccormick_editorial_2014}. The study of the role of neuromodulation in shaping rhythmic circuits, and in particular \glspl{cpg}, has provided fundamental insights into how neural systems achieve quick and robust adaptability to changing environments. This ability to flexibly switch between behaviors without structural rewiring of neuronal circuits can be a powerful, yet largely unexplored source of inspiration for neuromorphic hardware design~\cite{bartolozzi_embodied_2022}.

The opportunities provided by neuromodulation are illustrated in \Cref{fig:nmd_bio}. The top traces are physiological recordings of a neuron undergoing neuromodulation, switching between slow tonic spiking and transient bursting. The first mode supports linear rate-based encoding of the input strength, while the second allows a nonlinear detection of input changes, a "wake-up call" response. The bottom traces show the same modulation recorded experimentally on our neuromorphic chip presented in this work, where varying a single bias voltage enables reliable behavioral switches across all neurons present on chip. Capturing such transitions in silicon not only brings us closer to the richness of biological computation but also enables a new range of possibilities for flexible and adaptive behaviors in neuromorphic systems. This example is one of many possibilities for functional switches in behaviors achievable using neuromodulation.

\begin{figure}[!t]
    \centering
    \includegraphics{Figures/NMD_intro.tikz}
    \caption{Neuromodulation in thalamocortical relay neurons (top, adapted from~\cite{sherman_tonic_2001}) vs. neuromodulation recorded in our silicon neuron (bottom). For the same input current step (middle trace), neuromodulation enables a switch between tonic spiking (left) and a single transient burst (right), corresponding respectively to a linear rate-based encoding or a nonlinear "wake-up call" response to changes. In thalamocortical relay neurons, this response change is caused by hyperpolarization, which can be driven by neuromodulators. In our silicon neuron, the same change is obtained by varying a bias voltage. Subthreshold integration dynamics are also reproduced in our neuron, but are much faster than biological neurons due to the limited capacitor sizes.}
    \label{fig:nmd_bio}
\end{figure}

Neuromorphic engineering aims to replicate neural computation using analog, digital, or mixed-signal circuits that mimic the asynchronous, event-driven, and massively parallel architecture of the brain~\cite{mead_neuromorphic_1990,Indiveri25,chicca_neuromorphic_2014,schuman_survey_2017,neftci_data_2018,yang_neuromorphic_2020}. In the analog and mixed-signal domains, subthreshold transistor operation and current-mode designs have naturally emerged as powerful techniques for implementing power-efficient circuits emulating neuronal dynamics~\cite{poon_neuromorphic_2011,chicca_neuromorphic_2014}. Subthreshold circuits exploit the exponential current-voltage relationship of \glspl{fet} near the threshold voltage to generate nonlinearities at ultra-low currents, and current-mode design allows direct addition and subtraction of signals, across several orders of magnitude of current ranges despite limited voltage headrooms~\cite{mead_neuromorphic_1990}.
Most modern silicon neuron designs for large-scale analog and mixed-signal implementations are based on variations of compact \gls{iandf} models~\cite{thakur_large-scale_2018,Wunderlich_etal19}. Their efficiency, robustness, and compactness make them highly scalable but come at the cost of missing most of the key dynamical features needed to emulate biological neuromodulation in silicon. More than a biological curiosity, neuromodulation is a fundamental neural mechanism for generating diverse temporal patterns, enabling context-dependent information processing, and supporting the emergence and regulation of different behavioral states. The ability to flexibly switch between dynamical regimes is crucial for functions including rhythmic pattern generation, sensory filtering, and behavioral switching~\cite{avery_neuromodulatory_2017,mccormick_editorial_2014}. Reproducing it in compact, low-power neuromorphic hardware remains a challenge to date, as it involves using larger neuron models containing various nonlinear interactions between a larger number of state variables across multiple timescales. Some works have started addressing this issue~\cite{hynna_silicon_2007,yu_analog_2010,liu_bionn_2023}, but do not achieve the level of integration and power efficiency required by large-scale integration.

Nevertheless, both neuroscience and neuromorphic computing communities increasingly recognize the computational and functional benefits of neuromodulation and more complex excitable behaviors. Bursting in particular has been shown to enhance the performance of artificial spiking neural networks in learning tasks~\cite{li_bursting_2016,liu_artificial_2023,stuck_burst-dependent_2025,mei_improving_2025}. Translating the robust and tunable rhythmic properties of bursting \glspl{cpg} into hardware has also been the focus of several recent works~\cite{ijspeert_central_2008,donati_neuromorphic_2021,athota_neuromorphic_2021,lopez-osorio_neuromorphic_2022,liu_bionn_2023}. However, current large-scale analog and mixed-signal neuromorphic platforms, such as~\cite{pehle_brainscales-2_2022,aamir_accelerated_2018,richter_dynap-se2_2024}, lack native and robust support for intrinsic bursting and neuromodulation. Instead, these platforms rely on generating population-level emergent rhythms from large groups of spiking neurons or on exploiting biologically implausible dynamical features. Such strategies either lead to non-robust behaviors or resource-intensive implementations, which limits their applicability in real-world scenarios. The only exception to date is the silicon neuron presented in~\cite{liu_bionn_2023}. However, its voltage-mode design offers lower dynamic range and energy efficiency compared to current-mode subthreshold designs.

Recently, "mixed-feedback" modeling has emerged as a promising framework to address the trade-off between biological plausibility, behavioral richness, and hardware viability. In mixed-feedback models, the complex dynamics of numerous interacting neural ionic currents are captured by the superposition of a reduced number of positive and negative feedback loops on a well-defined hierarchy of timescales~\cite{franci2013balance,drion_dynamic_2015,castanos2017implementing,ribar_neuromodulation_2019,liu_bionn_2023,sepulchre2019control}. These models produce rich and robustly tunable neuronal behaviors while remaining low-dimensional and tractable. The minimality of mixed-feedback models paves the way toward plausible large-scale implementations of neuromodulable neurons and also makes them more tractable for systematic analysis and tuning, as already highlighted in~\cite{ribar_neuromodulation_2019}.\newline

In this work, we bridge the gap between the theoretical richness of mixed-feedback models and state-of-the-art analog neuromorphic circuit design. We present a fully analog mixed-feedback neuromodulable neuron implemented in current-mode subthreshold circuits~\cite{chicca_neuromorphic_2014,rubino2021ultralow,rubino2025Phd}. Our design preserves the core dynamical structure of mixed-feedback models and introduces a series of practical improvements tailored to reduce power consumption and area. The resulting circuit exhibits fully modulable excitable behaviors while remaining compact and low-power. This approach lays the foundation for scalable neuromorphic systems that can exploit circuit-level neuromodulation for adaptive behavior and context-aware computation in real-world scenarios.

In \Cref{sec:circuit}, we introduce the proposed mixed-feedback neuron model and describe its implementation in subthreshold current-mode \gls{cmos} circuits for a compact, modular, and ultra-low-power hardware realization of a neuromodulable neuron. In the model, the different timescales are continuously coupled, as opposed to being linked by events only. This subtle difference drastically enhances the expressivity of our model. We detail the use of \acrfull{dpi}~\cite{bartolozzi_ultra_2006} and current-mode sigmoid blocks, including a novel biologically inspired positive feedback inactivation mechanism. We also present an analysis framework that enables prediction and tuning of the neuron's excitable behavior by connecting circuit-level tuning and model equations. In \Cref{sec:results}, we present experimental results obtained from a \SI{180}{\nano\meter} \gls{cmos} prototype that exhibits consistent spiking, bursting, and neuromodulation capabilities. We further validate the circuit's resilience to temperature changes (\qtyrange{5}{45}{\celsius}) and report an energy efficiency, at room temperature and as a function of spiking frequency, of \qtyrange{40}{200}{\pico\joule}/spike at a \SI{1.8}{\volt} supply. We also evaluate robustness to device mismatch in simulation and show that tuning the characteristic scale of bias currents enables progressive tuning from an ultra-low-power mode to a strongly reliable mode, while maintaining subthreshold operation. In \Cref{sec:discussions}, we compare our design with state-of-the-art bursting neuron implementations and discuss integration within larger neuromorphic systems and platforms. Since this work focuses on the design, analysis, and experimental validation of a single neuron circuit rather than a complete neuromorphic system, we frame the proposed neuron as a modular building block that can be integrated into a wide range of architectures and communication schemes, independent of a specific system-level organization. Finally, \Cref{sec:concl} concludes the paper and outlines future research directions.
\section{The mixed-feedback neuron circuit}\label{sec:circuit}

To capture the essential dynamics of neuronal excitability and neuromodulation and implement them efficiently in neuromorphic hardware, we adopt a mixed-feedback approach~\cite{franci2013balance,drion_dynamic_2015,ribar_neuromodulation_2019,liu_bionn_2023,sepulchre2019control} that uses a superposition of positive and negative feedbacks across distinct timescales. In contrast to the few existing mixed-feedback neuron implementations~\cite{ribar_neuromodulation_2019,liu_bionn_2023}, we propose a mixed-feedback architecture using current-mode circuits, ensuring more compact, robust, and energy-efficient designs~\cite{chicca_neuromorphic_2014}. All state variables, such as the neuron's membrane "potential" and the positive and negative feedback terms, are therefore represented as physical currents. This choice naturally supports the modularity of the feedback design by exploiting Kirchhoff's current law to implement sum and difference operations. It also ensures compatibility with ultra-low-power subthreshold CMOS circuits, even in technology nodes characterized by limited voltage headroom~\cite{rubino2021ultralow,rubino2025Phd}.

This section introduces the structure of our mixed-feedback neuron model and its specificities, details its implementation using subthreshold current-mode CMOS circuits, and presents a mathematical formulation for the analysis and tuning of the circuit.

\subsection{A current-mode mixed-feedback neuron model} \label{subsec:modelintro}

\begin{figure}[!t]
    \centering
    \includegraphics{Figures/block_diagram.tikz}
    \caption{Mixed-feedback neuromodulable neuron structure. Red (\textit{resp.} blue) highlighted arrows indicate positive (\textit{resp.} negative) feedback paths. The two types of blocks use the notations introduced in \cref{eq: lpf state space,eq: sigmoid}: the low-pass filter blocks are denoted using their transfer function, and the sigmoid blocks using their symbol, their gain variable, and the shape of their steady-state response. The dashed arrows indicate positive feedback inactivation described by \cref{eqs:inact}, they represent a modulation of the indicated gain parameter of the sigmoids they traverse. $I_f$ is the current-mode analog of the membrane potential. $I_s$ and $I_u$ respectively provide slow repolarization and ultraslow spike-frequency adaptation currents. The nonlinear sigmoid blocks provide positive feedback that creates the fast spike upstroke and slow regenerative dynamics.}
    \label{fig:neuron_diagram}
\end{figure}

The design of a mixed-feedback architecture requires two building blocks: first-order low-pass filters and static sigmoidal functions.

A first order current-mode low-pass filter, with input $I_\text{in}$ and output $I_\text{out}$, is described in state-space representation as
\begin{align}\label{eq: lpf state space}
    \tau\frac{\mathrm{d}{I}_\text{out}}{\mathrm{d}t}&=-I_\text{out}+G I_\text{in}\,,
\end{align}
where $\tau$ is the time constant of the filter and $G$ its gain. In the Laplace domain, \eqref{eq: lpf state space} is represented by the transfer function
\begin{equation}
    \frac{I_\text{out}}{I_\text{in}}=H(s)=\frac{G}{\tau s+1}\,,
\end{equation}
with $s$ the complex variable of the Laplace domain. Low-pass filters serve two complementary purposes in the mixed-feedback architecture. First, they enable the accumulation and integration of signals over time. Second, they create well-separated time constants for different feedback pathways, reflecting the variety of timescales observed in neural ion channel kinetics~\cite{Hille2001}.

We denote static sigmoidal current-mode input-output characteristics as
\begin{equation}\label{eq: sigmoid}
    I_\text{out}=\mathcal S(I_\text{in};I_\text{G})\,, 
\end{equation}
where $I_\text{G}$ is the sigmoid gain. Such sigmoidal nonlinearities serve as a minimal model of neural ionic currents with gated and bounded conductances. These nonlinear functions introduce thresholding and saturation, which are key to creating the rich and tunable neuronal responses of mixed-feedback models. Examples of such current-mode sigmoidal functions will be shown in \cref{fig:sigmoid_response}. Importantly, the combination of saturating nonlinearities and separated time constants enables excitable feedback dynamics that are expressive, robust, and tunable through a small number of parameters, allowing the neuron to exhibit qualitatively distinct firing regimes under different neuromodulation settings.

Using these two building blocks, our mixed-feedback neuron combines nonlinear feedback pathways across multiple timescales into a compact dynamical system. The resulting structure, implemented exclusively with current-mode low-pass filters and sigmoidal nonlinearities, is represented in \cref{fig:neuron_diagram}. The internal dynamics of the model are organized around three sharply distinct timescales (fast, slow, and ultraslow) defined by the time constants of three low-pass filters, two linear negative feedback loops (highlighted in blue), and two nonlinear positive feedback loops (highlighted in orange). The circuit implementations of both types of blocks will be discussed in \cref{subsec:circuitblocks}.

The fast low-pass filter acts as a leaky integrator and models the passive membrane dynamics of the neuron. Its output, the current $I_f$, serves as a current-mode analog of the neuron's membrane potential. It receives the external inputs to the neuron $I_{app}$ and integrates the effects of all the feedback loops, including the positive self-feedback loop through $\mathcal S_{f}$. The two other low-pass filters define the slow and ultraslow current-mode state variables $I_s$ and $I_{u}$. $I_s$ provides both linear slow negative feedback and nonlinear slow positive feedback, through $\mathcal S_{s}$, on the membrane potential $I_f$, while $I_{u}$ provides ultra-slow negative feedback. These feedback pathways correspond to those provided by ionic current on membrane potential variations in biological neurons~\cite{franci2013balance,drion_dynamic_2015,hodgkin_quantitative_1952,izhikevich_dynamical_2006,Hille2001}: the fast positive feedback mimics the rapid activation of sodium channels responsible for spike initiation; the slow positive feedback simulates low-threshold calcium currents that support bursting or regenerative depolarization; the slow and ultraslow negative feedbacks capture delayed-rectifier and calcium-activated potassium currents responsible for spike repolarization, spike-frequency adaptation, and burst termination.

\begin{figure}[!t]
    \centering
   \includegraphics{Figures/inact.tikz}
    \caption{Spike (A) and burst (B) with (right) and without (left) positive feedback inactivation. The temporal extent of spikes and bursts is reduced thanks to the inactivation, but they remain long enough in time to directly signal to physical systems (in the spirit of \cite{deweerth_simple_1991}).}
    \label{fig:inactivation}
\end{figure}

Besides its current-mode nature, our neuron design departs from~\cite{ribar_neuromodulation_2019,liu_bionn_2023} in two other fundamental aspects. First, the negative feedback loops in the proposed mixed-feedback architecture are linear, removing additional gating and saturating sigmoidal blocks. This simplification reduces circuit complexity, area, and power consumption, while minimally affecting robustness and tunability. Indeed, negative feedback enforces stability and adaptation, whereas regime selection and neuromodulation are governed by the nonlinear positive feedback pathways. This design choice preserves the qualitative shape of the steady-state characteristics that determine firing regimes, as confirmed by the steady-state curves reported later in \cref{subsec:analysis}. Second, inspired by ionic current inactivation kinetics in biological neurons~\cite{hodgkin_quantitative_1952,Hille2001,zang_sodium_2023}, our model includes a positive feedback inactivation mechanism that promotes shorter and more power-efficient spikes.
In practice, the gains of the fast and slow positive feedbacks are modulated by subtracting the slow $I_s$ and ultraslow $I_u$ currents from the fast $I_{\text{G}_f,0}$ and slow $I_{\text{G}_s,0}$ sigmoid gains, respectively. The resulting modulated sigmoid gains $I_{\text{G},f}(t)$ and $I_{\text{G},f}(t)$ are defined as
\begin{subequations}\label{eqs:inact}
\begin{align}
   I_{\text{G}_f}(t)&=I_{\text{G}_f,0}-I_{s}(t)\,,\\
   I_{\text{G}_s}(t)&=I_{\text{G}_s,0}-I_{u}(t)\,.
\end{align}
\end{subequations}

Because the fast dynamics are assumed to be much faster than the slower ones, the multiplicative interactions created by positive feedback inactivation have a negligible impact on spike initiation but become significant during repolarization. By attenuating positive feedback right after a spike is generated, it shortens spike duration, lowering per-spike power consumption and making spike events more localized in time, without interfering with the underlying excitability and modulation properties. The effect of positive feedback inactivation is illustrated in \cref{fig:inactivation}, which compares simulations of spiking and bursting traces with and without inactivation.

These waveforms illustrate a full spike or burst cycle, including an input integration phase, a rapid depolarization driven by the nonlinear positive feedback, followed by repolarization mediated by the slower negative feedback pathways implemented through the low-pass filters. The fast timescale governs spike onset, while the slow and ultraslow dynamics progressively suppress the positive feedback, bringing the neuron back to its resting state and enforcing a refractory period. For bursting, the slow positive feedback enables new spikes after repolarization, before the ultraslow negative feedback suppresses the activity entirely, ending the burst.

\subsection{Implementation of current-mode circuit blocks}\label{subsec:circuitblocks}

To implement current-mode low-pass filters, we use the \gls{dpi} circuit~\cite{bartolozzi_ultra_2006} shown in \cref{fig:DPI_circuit}. The \gls{dpi} has become a staple in analog and mixed-signal neuromorphic systems due to its compact design, low-power operation, and tunable gain and timescale.

\begin{figure}[!t]
\centering
\includegraphics[width=0.6\singlefigwidth]{Figures/DPIblock.tikz}

\includegraphics{Figures/DPIcircuit.tikz}

\caption{\Acrfull{dpi} circuit (from~\cite{bartolozzi_ultra_2006,bartolozzi_synaptic_2007,livi_current-mode_2009,chicca_neuromorphic_2014}), tunable via its two bias voltages $V_\text{th}$ and $V_\tau$. Q2-3 are the differential pair receiving the input current $I_\text{in}$. With Q1, Q2 sets the gain $G=\frac{I_\text{th}}{I_\tau}$ of the circuit by diverting part of the input current depending on $I_\text{th}$, while Q3 sets the voltage $V_C$ of the capacitor. Q4 provides the leakage current $I_\tau$ that discharges the capacitor over time. Q5 provides an output current $I_\text{out}$ depending on the capacitor voltage $V_C$.}
\label{fig:DPI_circuit}
\end{figure}

Under the assumptions of subthreshold operation, and provided that the input current exceeds the leakage current ($I_\text{in} > I_\tau$), so that the output current will eventually dominate the threshold current ($I_\text{out}\gg I_\text{th}$), its dynamics can be approximated by those of a first-order low-pass filter, as expressed in~\eqref{eq: lpf state space}, with a gain $G=\frac{I_\text{th}}{I_\tau}$ and a time constant $\tau = \frac{C U_T}{\kappa I_\tau}$, where $I_\tau = I_0e^\frac{\kappa V_\tau}{U_T}$, $I_0$ is the transistor leakage current, $\kappa$ is the subthreshold slope factor, and $U_T$ is the thermal voltage~\cite{livi_current-mode_2009}. Both $G$ and $\tau$ are tunable via the two bias voltages $V_\text{th}$ and $V_\tau$.

The mixed-feedback model relies on a strong separation between the different timescales to justify the quasi steady-state approximation that supports its theoretical structure (\cref{subsec:analysis}). In hardware, this translates into \gls{dpi} filters with time constants spanning multiple orders of magnitude, implemented here through scaling of the capacitor sizes. In our circuit, we used capacitors ranging from hundreds of \si{\femto\farad} for the fast timescale to tens of \si{\pico\farad} for the ultraslow one, as reported in \cref{tab:dimensions}. Capacitor sizes increase by a factor of 4 between each timescale, setting the basis to achieve the desired timescale separation. Indeed, considering a leakage current $I_\tau=\SI{2}{\pico\ampere}$, which is achievable in mature \gls{cmos} nodes, one obtains $\tau_f=\SI{9}{\milli\second}$, $\tau_s=\SI{35}{\milli\second}$, and $\tau_u=\SI{143}{\milli\second}$ at room temperature using the expression for $\tau$ reported above. This range is consistent with the fast, slow, and ultraslow dynamics of spiking and bursting in cortical neurons~\cite{sherman_tonic_2001}, as illustrated in \cref{fig:nmd_bio}, and provides a sufficient timescale separation to ensure proper circuit operation. If desired, the leakage currents of the individual \glspl{dpi} may be further adjusted to accentuate the differences between timescales. As will be discussed later in \cref{subsec:comparison}, these capacitors occupy most of the neuron's area on chip, representing the main physical cost of faithfully implementing biologically plausible timescales.\newline

The two nonlinear sigmoids $\mathcal S_f(I_{f};I_{\text{G}_f})$ and $\mathcal S_s(I_{s};I_{\text{G}_s})$ in the neuron model are implemented using a custom current-mode subthreshold circuit, shown in \cref{fig:sigmoid_circuit}. This design combines a simple comparator with diode-connected transistors and a gain branch to produce a fully tunable sigmoidal current-to-current relationship. 
Three bias voltages (associated with bias currents through the standard subthreshold equation of NFETS in the saturation regime $I_b = I_0e^\frac{\kappa V_b}{U_T}$) independently provide full control over the sigmoid input-output characteristic. $V_\text{thr}$ sets the input threshold current $I_\text{thr}$ at which the circuit starts to output current; $V_\text{lin}$ controls the size of the input current range over which the sigmoidal output is monotonically increasing and roughly linear; and $V_\text{G}$ sets the maximum output current $I_\text{G}$. The effect of these three bias currents is illustrated in~\cref{fig:sigmoid_response}. The result is a compact, low-power, and easily tunable current-mode sigmoid function, well-suited for designing neuromorphic circuits with nonlinear feedback dynamics. Mathematical developments proving the sigmoidal shape of the input-output relationship of the circuit are shown in Appendix~\ref{sigmoidappendix}.

\begin{figure}[!t]
    \centering
   \includegraphics[width=0.6\singlefigwidth]{Figures/sigmoidblock.tikz}
   
   \includegraphics{Figures/sigmoidcircuit.tikz}
    
    \caption{Current-mode sigmoid circuit with inactivation mechanism. The three bias voltages $V_\text{thr}$, $V_\text{lin}$ and $V_\text{G,0}$ are assumed to be obtained through NMOS current mirrors and linked to the bias current parameters $I_\text{thr}$, $I_\text{lin}$ and $I_\text{G,0}$ through $I_b = I_0\exp\left(\frac{\kappa V_b}{U_T}\right)$. Q1-2 form a simple current comparator circuit: when $I_\text{in}$ > $I_\text{thr}$, the voltage $V_\text{cmp}$ starts to increase. Q3-5 control the rate of increase of $V_\text{cmp}$: the higher $V_\text{lin}$ is, the more current is drawn by this branch, increasing the response range of the comparator. Q6-7 set the gain of the circuit, defining the maximum output current $I_\text{G}$ at which the sigmoid saturates. The role of M1-5 is to implement the inactivation mechanism by subtracting $I_\text{mod}$ from the parameter $\hat{I}_\text{gain}$.}
    \label{fig:sigmoid_circuit}
\end{figure}

\begin{figure}[!t]
    \centering
    \includegraphics{Figures/sigmoid_sim.tikz}
    \caption{Current-mode sigmoid circuit response for different bias parameter values. The bias currents $I_\text{thr}$, $I_\text{lin}$, and $I_\text{G}$ accurately set the input threshold, the width of the monotonically increasing range, and the maximum output current respectively, for any combination of subthreshold bias parameters.}
    \label{fig:sigmoid_response}
\end{figure}

The circuit shown in~\cref{fig:sigmoid_circuit} also includes a gain modulation subcircuit that implements the positive feedback inactivation dynamics described in~\cref{subsec:modelintro}. The inactivation current $I_\text{mod}$ is subtracted from the baseline gain current $I_{\text{G},0}$ using current mirrors, such that the effective gain bias voltage $V_\text{G}$ fed to the sigmoid gain branch produces a gain current $I_\text{G}(t) = I_{\text{G},0} - I_\text{mod}(t)$.\newline

Using the \gls{dpi} and sigmoid circuits, the entire mixed-feedback model presented in~\cref{fig:neuron_diagram} can be implemented with only two different analog blocks, as shown in the full circuit schematic in~\cref{fig:fullschematic}, simplifying not only tuning but also chip design and layout. This modularity is a key strength of the approach: each feedback loop is implemented using the same circuit templates, with the resulting feedback behavior controlled entirely through the 12 bias currents (3 for each of the 2 sigmoids, and 2 for each of the 3 \glspl{dpi}). The use of subthreshold current-mode design further ensures ultra-low-power operation, making the neuron well-suited for large-scale deployment in power-constrained neuromorphic systems. Moreover, this neuron can easily be interfaced with modern event-based communication schemes such as \gls{aer} to enable large-scale communication. Input current can be directly provided by \gls{dpi} synapses, while spike events can be extracted from the fast sigmoidal output node $V_{\mathcal{S}_f}$ and converted into digital pulses, creating a natural separation between continuous-time internal dynamics and spike signaling.

\paragraph{Current-scale invariance}
    Because all state variables are currents and interact through linear \gls{dpi} filters and current-mode summation, the model dynamics are invariant under uniform scaling of the bias currents, provided the system operates in the subthreshold, or weak inversion, regime where the drain current exhibits an exponential dependence on gate voltage. This region is usually defined by $V_{GS}-V_{th}\leq \qtyrange{3}{4}{}\,U_T$, where the threshold voltage $V_{th}$ is technology-dependent~\cite{razavi_design_2017}. In this regime, a uniform scaling of all biases indeed preserves the relative strength of the positive and negative feedback pathways, as justified in~\cref{subsec:analysis}.
    
    In our \SI{180}{\nano\meter} technology, the neuron maintains its qualitative dynamical behavior up to bias currents of approximately \SI{7}{\micro\ampere} (for the fast sigmoid gain $I_{\text{G},0_f}$). Beyond this point, a departure from the predicted scale invariance is observed as devices enter moderate inversion and the subthreshold assumptions behind the derivation of the \gls{dpi} dynamics \eqref{eq: lpf state space}, which justify the operation of the mixed-feedback model in~\cref{subsec:analysis}, are no longer met. Increasing the operating current within the subthreshold limits reduces mismatch sensitivity (as shown later in \cref{subsubsec:mismatch}) at the expense of higher power consumption, while lower currents favor ultra-low-power operation. The lower bound for operating currents is thus set by the acceptable level of variability for a given application, down to typical transistor leakage currents of the technology. This flexibility makes the approach suitable for both energy-constrained systems and performance-critical scenarios.

\paragraph{Temperature invariance}
Similar arguments as for the current-scale invariance apply when considering temperature variations. Because the subthreshold currents scale with temperature, all feedback loop currents and time constants change proportionally, preserving their relative ratios. Consequently, the neuron dynamics remain qualitatively invariant, with temperature primarily rescaling the temporal evolution. This proportional scaling accounts for the temperature invariance experimentally observed in Figure~\ref{fig:temp_meas}.

\subsection{Model tuning and theoretical analysis}
\label{subsec:analysis}

To capture the continuous-time dynamics of our mixed-feedback neuron, we can leverage the structure of the model and write the input-output relationship of each low-pass filter, which yields
\begin{subequations}\label{eqs:neuronsystem}
    \begin{align}
            \tau_f \dot{I}_f =& - I_f + G_f\left(\mathcal{S}_f(I_f;I_{\text{G}_f}) + \mathcal{S}_s(I_s;I_{\text{G}_s})
            - I_s - I_{u} + I_{app}\right)\,,\label{eq:Ifast}          
        \\
            \tau_s \dot{I}_s =& - I_s + G_s I_f\,,\label{eq:Islow}\\
            \tau_{u} \dot{I}_u =& - I_{u} + G_{u} I_f\,.\label{eq:Iuslow}
    \end{align}
\end{subequations}

These equations enable the use of the techniques introduced in~\cite{franci2013balance,drion_dynamic_2015} for biophysical conductance-based models and in~\cite{ribar_neuromodulation_2019} for voltage-mode mixed-feedback circuits to analyze the emergence of spiking and bursting behaviors from the interaction of the different feedback loops in this continuous mixed-feedback model. They provide a fast and systematic way of tuning the various parameters of the model by revealing the excitability thresholds and bistable regions , allowing to skip lengthy transient simulations and tune bias voltages solely based on DC characteristics of the feedback loops.

To illustrate these techniques, we compute the steady-state "output-input" curves used in~\cite{ribar_neuromodulation_2019, liu_bionn_2023} of the three feedback loops. These curves are conceptually similar to the I-V (current-voltage) curves measured using voltage clamp experiments in neuroscience. To that end, we assume timescale separation, that is, $\tau_f \ll \tau_s \ll \tau_u$, and neglect the effect of positive feedback inactivation. This is a reasonable assumption since the fast (\textit{resp.} slow) inactivation depends on the slow (\textit{resp.} ultraslow) variable, which is negligible in the timescale of interest through a quasi steady-state approximation enabled by the different capacitor sizes described in \cref{subsec:circuitblocks}. We also set $G_f=G_s=G_u=1$, so that the \gls{dpi} circuits behave like linear leaky integrators. We can derive the expression of the three steady-state characteristics by sequentially setting the derivatives in \cref{eqs:neuronsystem} to zero, starting from the fastest subsystem and neglecting the effect of slower variables, to obtain
\begin{subequations}\label{eqs:ivcurves_simpl}
\begin{align}
    \dot{I}_f = 0 &\iff I_{app} = \bar{I}_f - \mathcal{S}f(\bar{I}_f; \bar{I}_{\text{G}_f})\,,\label{eq:ivfast}\\
    \dot{I}_f, \dot{I}_s = 0 &\iff I_{app} = 2\bar{I}_s - \mathcal{S}_f(\bar{I}_s; \bar{I}_{\text{G}_f}) - \mathcal{S}_s(\bar{I}_s; \bar{I}_{\text{G}_s})\,,\label{eq:ivslow}\\
    \dot{I}_f, \dot{I}_s, \dot{I}_u = 0 &\iff I_{app} = 3\bar{I}_u - \mathcal{S}_f(\bar{I}_u; \bar{I}_{\text{G}_f}) - \mathcal{S}_s(\bar{I}_u; \bar{I}_{\text{G}_s})\,.\label{eq:ivuslow}
\end{align}
\end{subequations}

\begin{figure}[!t]
    \centering
    \includegraphics{Figures/IVcurves.tikz}
    \caption{Examples of steady-state curves (left) of our mixed-feedback model and the corresponding neuron responses (right), using the parameters of \cref{tab:biases}. The red (\textit{resp.} cyan) regions are the bistability region of the fast (\textit{resp.} slow) subsystems. When the slow bistability region is entirely contained within the fast one (top), the neuron can only exhibit tonic spiking. When the slow bistability region threshold comes before the fast one (bottom), the neuron becomes capable of bursting. The y-axis of the ultraslow curve indicates which input current will put the neuron in a certain region.}
    \label{fig:IV_curves}
\end{figure}

Examples of these curves and the resulting simulated neuron activity are shown in \cref{fig:IV_curves}. The slope of these curves indicates the stability of the corresponding equilibria: regions of positive (negative) slope correspond to stable (unstable) equilibria. By comparing the relative positions of the fast, slow, and ultraslow bistability thresholds (corresponding to the maxima and minima), one can immediately infer whether the neuron will exhibit tonic spiking, bursting, or purely passive behavior. A region of negative slope in the fast steady-state curve translates to the existence of positive feedback, responsible for spike initiation. If a bistability threshold appears in the slow steady-state curve before the fast one, it shows a bistability between tonic spiking and silence, from which bursting can arise when an ultraslow variable slowly drives the system across this bistable region. A monotonically increasing ultraslow steady-state curve guarantees that, at the slowest timescale, the system is monotonically stable, ensuring that it will go back to rest after a spike or a burst~\cite{ribar_neuromodulation_2019}.

This geometric interpretation provides a powerful visual tool for understanding how feedback interactions and timescale separation generate complex dynamics. Moreover, these steady-state curves only depend on $\mathcal{S}_f$ and $\mathcal{S}_s$ and can therefore be analyzed by simply simulating the DC characteristics of the two sigmoid circuits.

This analysis also predicts the scaling properties of the model.
Because all state variables and parameters are currents and interact through linear current-mode filters, the relative threshold positions remain invariant under uniform scaling of the bias currents, explaining the current and temperature invariance discussed in \cref{subsec:circuitblocks}. These invariance properties, intrinsic to the mixed-feedback current-mode formulation, contribute to the robustness of the proposed neuromorphic architecture.
\section{Experimental results}\label{sec:results}

To validate the proposed mixed-feedback neuron architecture and its current-mode implementation, we fabricated and experimentally characterized the silicon neuron under various operating conditions. This section demonstrates that the fabricated circuit reproduces the key behaviors predicted by the model, including spiking and bursting dynamics, robust neuromodulation, low-power operation, and temperature invariance. We first describe the experimental setup and measurement methodology, followed by an analysis of the basic excitability and modulation properties. Specifically, we demonstrate the emergence of bursting through tuning of a single parameter and highlight the relationship between input strength and firing frequency, both for slow (Type-I) tonic spiking and tonic bursting. We then report power consumption simulations for different firing patterns and frequencies. We also provide Monte Carlo mismatch simulations, showing a tradeoff between energy efficiency and robustness comparable to the state-of-the-art. Finally, we present the results of a temperature variation experiment, showing how our mixed-feedback neuron behaves at various operating temperatures. All measurements focus on neuron-level behavior and metrics, independently of a specific neuromorphic system or application.

\subsection{Experimental setup}\label{subsec:expsetup}

\begin{figure}[!t]
    \centering
  \begin{tikzpicture}
      \node[anchor=south west, inner sep=0] (image) at (0,0) {\includegraphics[width=6.7cm]{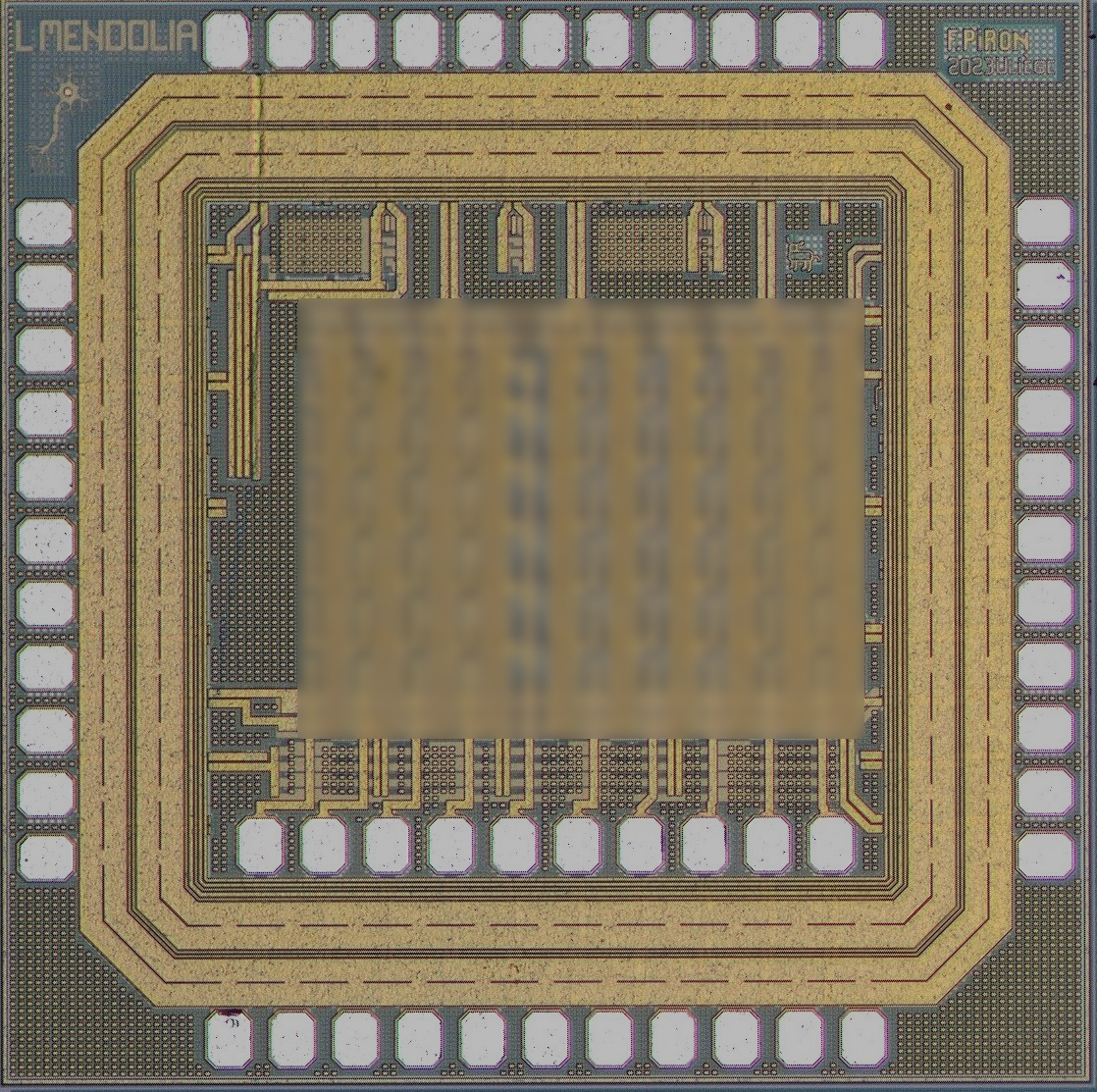}};
      \begin{scope}[x={(image.south east)},y={(image.north west)}]
          \draw[blue,ultra thick,rounded corners] (0.25,0.26) rectangle (0.77,0.33) node[midway, above=4pt, font=\tiny\bfseries] {Neuron array};
          \draw[green,ultra thick,rounded corners] (0.185,0.34) rectangle (0.22,0.38) node[above, inner sep=1pt, align=left, font=\tiny\bfseries] {Out.\\[1pt]buffer};;
          \draw[red,ultra thick,rounded corners] (0.2,0.55) rectangle (0.24,0.75) node[left, inner sep=1pt, align=left, font=\tiny\bfseries, yshift=-5pt, xshift=-7pt] {Shift\\[1pt]regs.};
          
          \node[anchor=south west, inner sep=0] (inset) at (0.35,0.4) {\includegraphics[width=2.68cm, trim={0cm 0cm 18cm 10cm},clip]{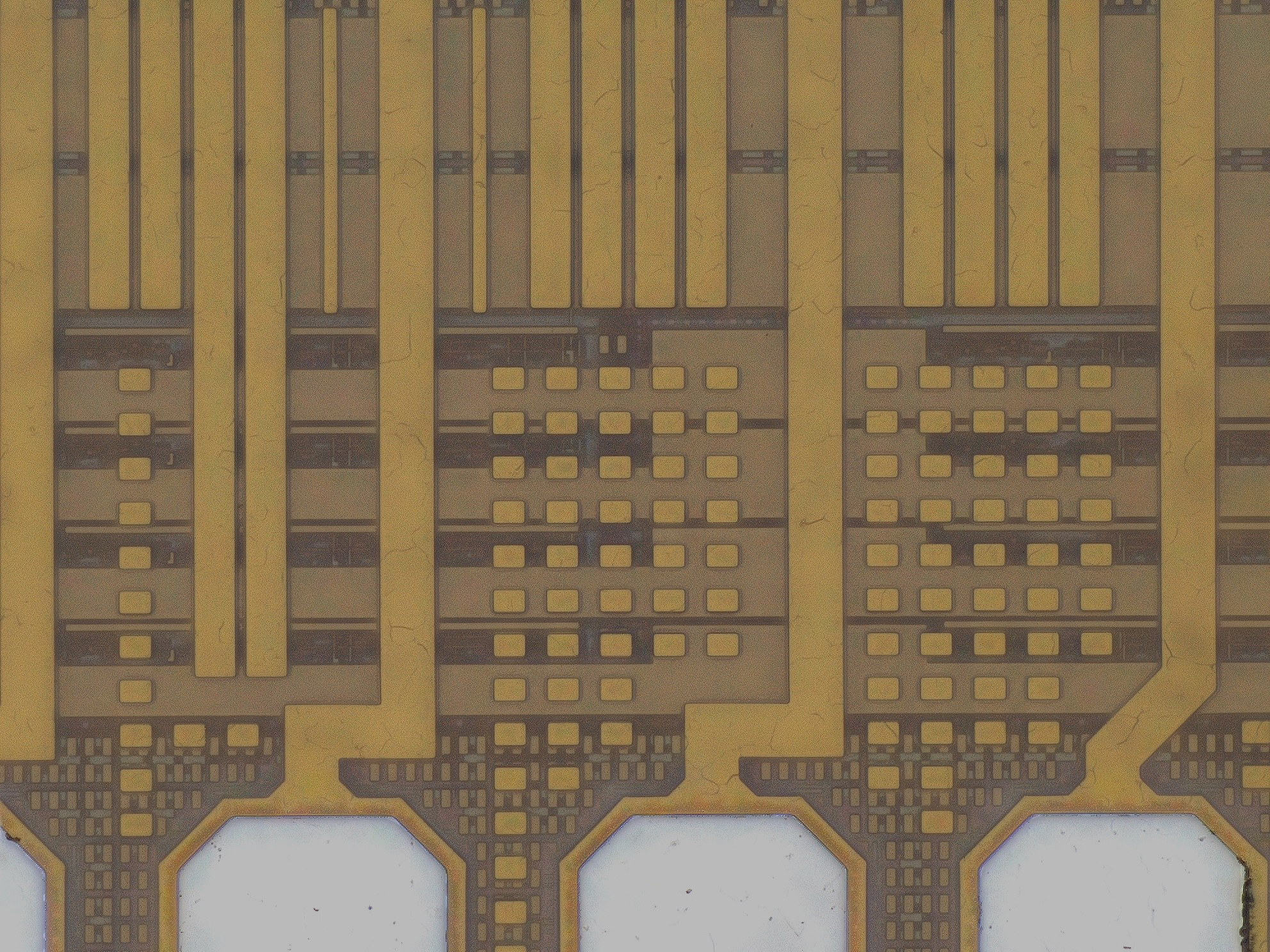}};

          \draw[white,line width=2.5pt, rounded corners=5pt] 
              (inset.south west) rectangle (inset.north east);

          \draw[white,ultra thick,dashed] (inset.south west) -- (0.42,0.29);
          \draw[white,ultra thick,dashed] (inset.south east) -- (0.54,0.29);

          \draw[white, ultra thick] (0.39, 0.5) -- ++(0.06, 0) node[midway, white, below]{\scriptsize \SI{30}{\micro\meter}};

          \draw[yellow, ultra thick, rounded corners=5pt] (0.375, 0.65) rectangle (0.73, 0.705) node[midway, above=2.3pt, font=\tiny\bfseries] {Single neuron};

          \draw[white, ultra thick] (0.02, 0.06) -- ++(0.12, 0) node[midway, white, below]{\scriptsize \SI{200}{\micro\meter}};
      \end{scope}
  \end{tikzpicture}
    \caption{Micrograph of the prototype chip (2023) containing the test mixed-feedback neurons. Blue: array of the 16 mixed-feedback neurons. Green: output buffer stage. Red: shift registers. Inset: zoom on 4 mixed-feedback neurons stacked on top of one another. Yellow: a single neuron. Circuits unrelated to this work are blurred.}
    \label{fig:chip_photo}
\end{figure}
\begin{figure}[!t]
    \centering
    \begin{tikzpicture}
    \node[anchor=south west, inner sep=0] (image) at (0,0){\includegraphics[width=6.7cm]{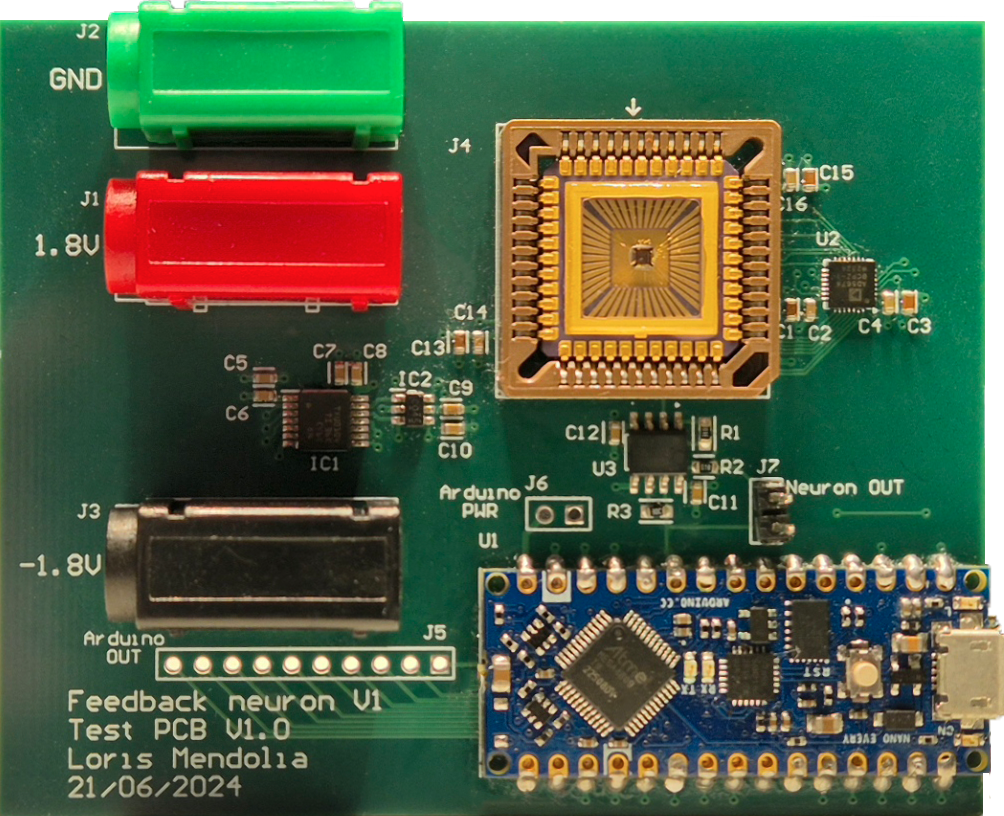}};
    \begin{scope}[x={(image.south east)},y={(image.north west)}]
    \draw[yellow,ultra thick,rounded corners] (0.5,0.51) rectangle (0.78,0.85) node[midway, above=27pt, font=\tiny\bfseries] {Socketed chip};
    \draw[lightgray,ultra thick,rounded corners] (0.25,0.42) rectangle (0.44,0.55) node[midway, left, align=right, font=\tiny\bfseries, xshift=-17pt] {Digital level\\[1pt]shifters};
    \draw[cyan,ultra thick,rounded corners] (0.57,0.38) rectangle (0.73,0.5) node[midway, right, align=left, font=\tiny\bfseries, xshift=15pt, yshift=3pt] {Feedback\\[1pt]ammeter};
    \draw[red,ultra thick,rounded corners] (0.81,0.605) rectangle (0.882,0.695) node[midway, right, align=left, font=\tiny\bfseries, xshift=5pt, yshift=3pt] {\acrshort{dac}};
    \draw[green,ultra thick,rounded corners, overlay] (0.47,0.04) rectangle (1,0.32) node[midway, above=20pt, font=\tiny\bfseries, overlay] {Arduino};
    
    \end{scope}
    \end{tikzpicture}
    \caption{Custom \gls{pcb} used for characterization of the on-chip mixed-feedback neurons. The board hosts the fabricated CMOS die (center, yellow), an external 12-bit \acrshort{dac} (AD5674, red) for bias generation, and an operational amplifier configured as a feedback ammeter (blue) for on-board current-to-voltage conversion. A microcontroller (Arduino Nano Every, green) controls the on-chip multiplexer for output selection and programming of the \gls{dac} for bias generation. Output signals were acquired with an oscilloscope (Analog Discovery 3) for data acquisition.}
    \label{fig:PCB}
\end{figure}

An array of 16 mixed-feedback neurons was fabricated on a test chip shared among different projects in a \SI{180}{\nano\meter} CMOS X-FAB process. An active area of \qtyproduct[product-units=single]{750x90}{\micro\meter\squared} was dedicated to this end. A micrograph of the fabricated chip is shown in \cref{fig:chip_photo}, highlighting the neuron array, control logic, and output. All neurons share a common set of 12 bias voltages (for the 2 sigmoids and 3 \glspl{dpi}), generated by an external \gls{dac} (an Analog Devices AD5674 12-bit, 16-channel \acrfull{dac}). A single analog output is provided through a transmission gate multiplexer, controlled by on-chip shift registers. The transmission gates allow the connection of one of the three internal currents ($I_f$, $I_s$, or $I_u$) of one neuron at a time to an output buffer stage, injecting an amplified analog copy of the selected current directly to the output pad, ensuring a reliable off-chip readout of the \si{\pico\ampere} or \si{\nano\ampere} internal currents.

The output current is converted into a voltage on the test \acrfull{pcb} shown in \cref{fig:PCB} using a precision operational amplifier in feedback ammeter configuration. This configuration allows to easily achieve different gains as required. After a second amplification stage using an inverting amplifier, the resulting voltage is measured by a digital oscilloscope (Digilent Analog Discovery 3, 14-bit resolution at up to \SI{125}{\mega\sample/\second}, \SI{30}{\mega\hertz} bandwidth), enabling a highly accurate acquisition of the neural traces. An Arduino Nano Every is mounted on the \gls{pcb} to configure the shift registers and the \gls{dac}, and can optionally stream output data. All results presented here were acquired using the oscilloscope for improved accuracy and flexibility. Unless otherwise stated, all measurements were performed at room temperature and at the nominal supply voltage of \SI{1.8}{\volt}.

\subsection{Measured dynamics in silicon}\label{subsec:measurements}
To validate and characterize the behavior of our neuron in silicon, we performed various measurements on the fabricated test chip. These experiments aim to confirm the neuron's ability to robustly produce and switch between different firing regimes as predicted by the model's mixed-feedback theoretical analysis. Extensive experimental exploration was performed, with several experimental runs on different days, different neurons, and different bias baselines to ensure the consistency of qualitative results. In addition, simulated power consumption metrics are reported to benchmark energy per spike in spiking and bursting regimes and to quantify the baseline power draw at rest. The experimental results presented in this section focus on the qualitative behavior of individual neurons. As transistor transistor sizing and layout were not optimized in this first prototype, all neurons exhibit similar dynamical behaviors, but at neuron-specific bias conditions. As a result, quantitative assessment of mismatch across neurons under identical biasing is not meaningful in this implementation. Robustness to mismatch is therefore evaluated through Monte Carlo simulations based on a revised design incorporating mismatch-aware sizing and layout, reflecting the expected variability of an optimized implementation.

\subsubsection{Response to exogenous input currents}

\begin{figure}[!t]
    \centering
    \includegraphics{Figures/meas_stim.tikz}
    \caption{Measured neuron responses to increasing input current (bottom), in spiking (top) and bursting (middle) configurations. The neuron exhibits tonic excitability, showing firing frequencies increasing with input current in both modes, starting as low as a few \si{\hertz}.}
    \label{fig:stim_meas}
\end{figure}

\Cref{fig:stim_meas} shows the response of an on-chip neuron to an increasing input current. Both in tonic spiking (top) and bursting (middle) regimes, the neuron exhibits a consistent increase in firing activity with the input, as expected from the mixed-feedback model. Despite the different internal dynamics, in both configurations the neuron responds in a consistent and reproducible way. In line with biological neurons, the spiking and bursting frequency of our neuron increases in response to an increasing exogenous input. In bursting regime and for sufficiently large applied currents, the bursts eventually overlap, and the neuron transitions from bursting to fast spiking, once again in line with biological neurons.

\subsubsection{Response to neuromodulatory inputs}

\Cref{fig:NMD_meas} demonstrates the neuromodulation capabilities of the fabricated neuron by varying the slow positive feedback gain $I_{G_s}$ while keeping the input current constant. As the neuromodulation parameter is incrementally increased, the neuron smoothly transitions from a tonic spiking to a tonic bursting regime, with progressively longer burst durations and more spikes per burst. These measurements illustrate how neuromodulatory tuning of a single bias current enables a smooth and predictable transition between firing regimes. Bursting characteristics, such as interburst period or spikes per burst, are directly and predictably modulated through $I_{G_s}$. 

The observed neuromodulation dynamics are consistent with the theory of slow positive feedback: increasing the slow gain increases the slow current's contribution to spike and burst excitability without interfering with the fast positive feedback responsible for the spike generation itself. The neuromodulatory transition achieved here is analogous to those observed in biophysical conductance models by increasing the maximal conductance of slow regenerative ionic currents, which create slow positive feedback on membrane potential variations~\cite{franci2013balance,drion_dynamic_2015,franci_robust_2018}.

\begin{figure}[!t]
    \centering
    \includegraphics{Figures/meas_nmd.tikz}
    \caption{Measured neuromodulation from spiking to bursting (top) through an increase in slow positive feedback $I_{\text{G}_s,0}$ (bottom). The neuron undergoes a smooth transition from spiking to bursting, with a gradual increase in the number of spikes per burst and a consistent inter-burst interval.}
    \label{fig:NMD_meas}
\end{figure}

Together, the measurements presented in \cref{fig:stim_meas,fig:NMD_meas} establish the neuron’s ability to reliably operate in both tonic spiking and bursting modes, with tunable transitions between the two. Our hardware neuron exhibits tunable excitability across a wide input dynamic range, with interpretable and reliable spiking mode control via specific bias currents. The control of bursting through a single (slow positive feedback) bias illustrated here is only one out of the many neuromodulatory pathways that can be explored and achieved in the proposed current-mode neuromorphic design.\newline

As already remarked in \cref{subsec:analysis}, the neuron dynamics are preserved under a uniform scaling of all bias currents. This is a direct consequence of the current-mode mixed-feedback implementation: as long as the system remains in subthreshold and steady-state relationships are preserved, all time constants and feedback loops scale proportionally with bias currents. As a result, operating the neuron at \SI{100}{pA} or \SI{100}{nA} yields the same qualitative behaviors, from tonic spiking to bursting, with a tunable trade-off between energy efficiency and mismatch sensitivity.

\subsubsection{Simulated power consumption}\label{subsubsec:power}

SPICE simulations were used to estimate the power consumption of our neuron (based on a revised layout and transistor sizing reported in \cref{tab:dimensions} optimized for mismatch, relative to the version implemented in the first test chip). The reported metrics exclude bias generation, digital control, and output circuitry, and focus solely on the neuron circuit described in \cref{sec:circuit}. Due to the shared nature of the test chip presented in the rest of this section, experimental validation of the metrics reported here is not possible, but will be the subject of future works.
Under the same bias conditions used throughout the measurements in this section, a neuron consumes approximately \SI{3}{\nano\watt} at rest (zero input), with instantaneous power momentarily rising to \SI{10}{\nano\watt} during spike generation.

In Type-I tonic spiking at \SI{23}{\hertz}, the dissipation over a half-second window of continuous analog activity, including input integration, spike generation, frequency adaptation, and other subthreshold dynamics, is \SI{217}{\pico\joule}/spike. After modulation to reach tonic bursting with 3 spikes per burst at 18 bursts per second, the average dissipation drops to \SI{90}{\pico\joule}/spike. As the tonic spiking frequency is increased, the dissipation drops down to \SI{101}{\pico\joule}/spike at \SI{50}{\hertz}, and \SI{41}{\pico\joule}/spike at \SI{160}{\hertz}. These results highlight the neuron's low-power operation across multiple dynamical states, with increasing energy efficiency at higher spiking rates. As these energy per spike metrics pertain to complex, continuous-time, yet temporally sparse fully analog events, they must be distinguished from the fundamentally different dense, clock-driven "per MAC (Multiply-Accumulate)" benchmarks used in the digital accelerator hardware literature, where energy consumptions of $<\SI{1}{\pico\joule}$/MAC are frequently reported~\cite{gomony_achieving_2024}.

\subsubsection{Statistical variability and mismatch}\label{subsubsec:mismatch}
\begin{figure}[!t]
    \centering
    \includegraphics[scale=1]{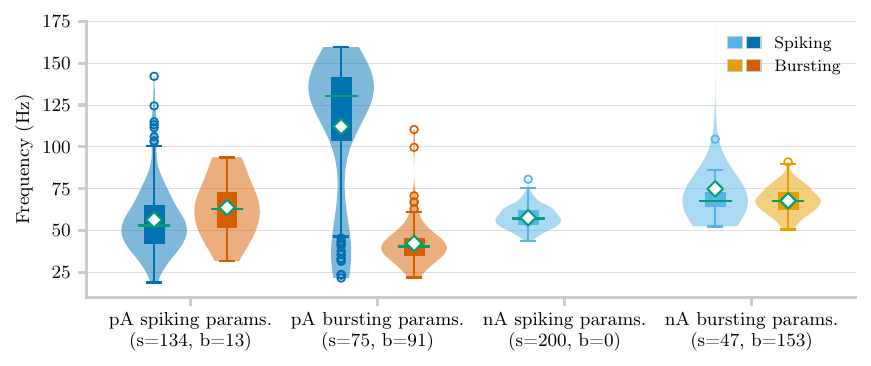}
    \caption{Monte Carlo mismatch analysis ($N=200$) results of nominal spiking and bursting configurations biased in the picoampere range (\cref{tab:biases}) and after $1000\times$ uniform bias scaling to the nanoampere range. Violin plots depict the kernel density of the frequency distribution for spiking and bursting neurons. The green bar marks the median, the diamond indicates the mean, the box spans the \acrfull{iqr} (Q1-Q3), and the whiskers extend to $1.5\times\text{\gls{iqr}}$, setting the bounds for outliers shown as hollow circles. At ultra-low currents, mismatch leads to regime mixing and larger frequency variability. Increasing bias currents restores full yield and reduces frequency dispersion.}
    \label{fig:mismatch}
\end{figure}

The prototype presented in this work focused on validating the mixed-feedback architecture and its dynamical properties, rather than optimizing device-level robustness to mismatch. In a subsequent design iteration, transistor sizing was optimized through statistical simulations, and systematic mismatch-aware layout techniques, such as common-centroid placement and use of dummy transistors, were incorporated. Transistor sizings are reported in Appendix~\ref{circuitappendix} and \cref{tab:dimensions}. To assess the expected variability in the optimized design, we performed Monte Carlo mismatch simulations of the neuron using foundry-provided models with 200 points under constant input for nominal spiking and bursting parameter sets in the picoampere regime (reported in \cref{tab:biases}). We then repeated the same simulations after uniformly scaling all bias currents by $1000\times$ into the nanoampere range and adjusting the input current to obtain comparable firing frequencies. \cref{fig:mismatch} shows the resulting distributions of spiking and bursting (interburst) frequencies (measured as the inverse of the period between the start of two bursts).

In the picoampere regime, the yield (number of neurons tonic firing) is between 75 and 80\%. For nominal spiking parameters, a few neurons transition to bursting, and firing frequency exhibits a \gls{cv} of 36\%. For nominal bursting parameters, about half of the neurons maintain bursting behavior, while the remaining active neurons predominantly transition to high frequency spiking. In that case, the \gls{cv} of firing frequencies is 31\% for bursting neurons and 38\% for spiking ones.

Scaling up the bias currents to the nanoampere range drastically improves reliability to mismatch. Yield improves to 100\% in both cases, and, in the spiking case, all neurons remain in the correct regime with a frequency dispersion of 11\%. Although 23\% of bursting neurons transition to slow tonic spiking under mismatch, average spiking and interburst frequencies are comparable and exhibit a dispersion of 11\% in bursting and 9\% in spiking (excluding 3 outliers), showing that the underlying timescale structure is better preserved. These figures are close to state-of-the-art current-mode neurons such as~\cite{rubino2021ultralow}.

These results reveal a clear tradeoff between energy efficiency and robustness to device mismatch. Ultra-low current operation maximizes energy efficiency but increases sensitivity to device variability, whereas current scaling, enabled by the current scale invariance property (\cref{subsec:circuitblocks}), improves regime stability and frequency reproducibility at the cost of increased power consumption. A sensitivity analysis of the simulations shows that the current mirrors formed by M$_f$1 and S$_f$7 (modulated fast positive feedback gain), Q4 and D$_u$4 (ultraslow negative feedback), and transistor S$_s$2 (slow positive feedback threshold) have the largest mismatch contribution to frequency variation and require the most attention during layout.

\subsection{Stability of the mixed-feedback model over time and temperature}\label{subsec:temperature}

To assess the robustness of our mixed-feedback neuron to environmental variations, we characterized the test chip's response across a wide temperature range using a climatic chamber. Proper measures were taken to ensure temperature stabilization with a thermal camera and to stay within the operating temperature range of the \gls{dac} used to generate bias voltages, also monitored throughout the experiment to ensure they remained constant.

\cref{fig:temp_meas} reports the evolution of the spiking and bursting (interburst) frequency of two representative neurons with temperature. In both cases, frequency increases gradually with temperature, but the overall spiking and bursting behaviors remain stable and qualitatively unchanged. The neuron responses at higher temperatures effectively correspond to sped-up or time-compressed versions of those at lower temperatures, a behavior analogous to the temperature robustness observed in biological neurons of species that experience large temperature variations, such as crabs~\cite{tang_robustness_2012,ratliff_neuronal_2021}. This observation further emphasizes the biophysical realism of our proposed mixed-feedback design.

Quantitatively, the measured spiking and inter-burst frequency increased from approximately \SI{0.5}{\hertz} at \SI{5}{\celsius} to \SI{25}{\hertz} at \SI{45}{\celsius}. These changes remained smooth and monotonic, without any loss of excitatory behavior or irregular activity. Across multiple trials and bias conditions, no degradation or drift was observed over repeated heating–cooling cycles, confirming the long-term stability and repeatability of the mixed-feedback neuron dynamics.

\begin{figure}[!t]
    \centering
    \includegraphics{Figures/temp_measurements.tikz}
    \caption{Measured variation in spiking (top) and bursting frequency (inverse of the period between the start of two bursts, bottom) of two different neurons as a function of ambient temperature, with representative output traces at two extreme temperatures. Numbers above the bursting data points indicate the number of spikes per burst. Frequency increases with temperature and follows a comparable trajectory across neurons and firing regimes. The recorded waveforms compress in time and speed up but remain in the same qualitative firing regime, confirming temperature-robust dynamics. The bursting neuron goes from 4-spike bursts at \SI{5}{\celsius} to 2-spike bursts at \SI{45}{\celsius} due to the accelerated ultraslow negative feedback dynamics.}
    \label{fig:temp_meas}
\end{figure}

The observed increase in frequency arises naturally from transistor-level temperature effects. As temperature increases, the thermal voltage $U_T = kT/q$ and the carrier mobility $\mu$ both change, resulting in higher subthreshold currents for identical gate bias conditions. In the \gls{dpi} filters, this results in proportionally faster charging and discharging, effectively reducing the time constants of the fast, slow, and ultraslow timescales. Otherwise, since all bias voltages are set by the same external source, changes in currents scale proportionally across all feedback loops, preserving the relative strength of the fast, slow, and ultraslow feedback pathways. As a result, the mixed-feedback neuron maintains its excitability regime, while its temporal dynamics accelerate uniformly. Note that, in the bursting regime, the number of spikes per burst decreases at higher temperatures. This is consistent with our model, since the ultraslow negative feedback accelerates with temperature without an accompanying positive component to counterbalance it. Importantly, the interplay between the fast and slow timescales which gives rise to bursting remains unaffected, and the neuron maintains its excitability type despite the change in burst length. Indeed, this interplay is determined by the relative organization of the bistability thresholds of the steady-state characteristics, as explained in \cref{subsec:analysis}, which is preserved under temperature variations, consistently with the scaling of all currents as per established analytical models of subthreshold operation~\cite{benjamin_analytical_2023}.

Overall, the mixed-feedback architecture provides an intrinsic form of temperature and bias self-regulation: when environmental conditions shift, the system’s internal currents scale coherently, preserving its qualitative behavior without requiring external compensation. This property further enhances the stability and robustness of the proposed model.
\newpage
\section{Discussion}\label{sec:discussions}

\subsection{Comparison of state-of-the-art bursting silicon neurons}\label{subsec:comparison}

To position our silicon neuron implementation alongside the state-of-the-art, we present a comparative overview of silicon neurons capable of generating bursting dynamics in the literature in \Cref{tab:comparison}. Unlike most prior work, which predominantly relies on voltage-mode implementations and simplified models such as the Izhikevich or AdEx neuron, our design offers a fully analog, current-mode implementation of the mixed-feedback model, extended here with a biologically motivated inactivation mechanism. This current-mode approach enables lower power consumption without sacrificing the richness of the dynamics, providing an efficient alternative to both digital implementations and transconductance-based analog designs. Compared to earlier works, such as those of Hynna \cite{hynna_silicon_2007} and Wijekoon \cite{wijekoon_compact_2008}, our neuron supports more diverse and biologically plausible behaviors while maintaining comparable or superior energy efficiency. The mixed-feedback model used here has been shown to uniquely exhibit a biologically consistent phase-plane structure, unlike the Izhikevich model, which suffers from limited tunability \cite{drion_novel_2012,schuman_survey_2017,jacquerie_robust_2021}.

\begin{table}[!t]
    \centering
    \small
    \label{tab:comparison}
    \begin{tblr}{width=\textwidth,colspec={l|X|X|X|X|l|X}}
        Ref. & Tech. & Impl. & Model & Biologically plausible NMD & Area & Energy/Power \\
        \hline
        \cite{hynna_silicon_2007} & \SI{0.25}{\micro\meter} CMOS & V-mode & T-Channel Axon-Hillock & Yes & \SI{774}{\micro\meter\squared} & ? \\
        \cite{wijekoon_compact_2008} & \SI{0.35}{\micro\meter} CMOS & V-mode & Izhikevich (accelerated) & No & \SI{2800}{\micro\meter\squared} & \SI{9}{\pico\joule}/spike (\qtyrange{8}{40}{\micro\watt}) \\
        \cite{livi_current-mode_2009} & \SI{0.35}{\micro\meter} CMOS & I-mode & AdEx & N/A & \SI{913}{\micro\meter\squared} & \SI{7}{\pico\joule}/pulse, \SI{267}{\pico\joule}/spike \\
        \cite{yu_biophysical_2011} & \SI{0.5}{\micro\meter} CMOS & V-mode & Extended HH \& Morris- Lecar & Yes & \SI{0.72}{\milli\meter\squared} & \SI{80}{\nano\joule}/spike (\SI{160}{\micro\watt}) \\
        \cite{folowosele_silicon_2011} & \SI{0.5}{\micro\meter} CMOS & Switched capacitors & Mihala\c{s}– Niebur & No & \SI{0.05}{\milli\meter\squared} & \SI{2.375}{\micro\watt} \\
        \cite{wang_compact_2015} & \SI{130}{\nano\meter} CMOS & Mixed & \gls{iandf} & No & \SI{26.5}{\micro\meter\squared} & High (FPGA) \\
        \cite{aamir_accelerated_2018, pehle_brainscales-2_2022} & \SI{65}{\nano\meter} CMOS & V-mode & AdEx (accelerated) & N/A & \SI{2352}{\micro\meter\squared} & \SI{790(170)}{\pico\joule}/spike \\
        \cite{liu_artificial_2023} & Pt/$\mathrm{Co}_3 \mathrm{O}_{4-x}$ /ITO & Memristive & -- & N/A & -- & -- \\
        \cite{liu_bionn_2023} & \SI{180}{\nano\meter} CMOS & V-mode & Mixed-feedback & Yes & \SI{2590}{\micro\meter\squared} &  \qtyrange{697}{969}{\nano\joule}/spike \\
        \cite{xiao_bio-plausible_2025} & $\mathrm{NbO}_2$ + ECRAM & Memristive & Fast-slow memristive & N/A & -- & -- \\
        \hline
        Our work & \SI{180}{\nano\meter} CMOS & Analog I-mode & Mixed-feedback (w/ inactivation) & Yes & \SI{3800}{\micro\meter\squared} & 40 to 200+ \si{\pico\joule}/spike \\
    \end{tblr}
    \caption{Comparison of state-of-the-art bursting silicon neurons (I\&F: integrate-and-fire, HH: Hodgkin–Huxley, AdEx: Adaptive Exponential I\&F).}
\end{table}

While recent work by Liu et al. \cite{liu_bionn_2023} has demonstrated a first silicon implementation of the mixed-feedback model in voltage-mode, our current-mode architecture further improves energy efficiency (down to \SI{45}{\pico\joule}/spike at \SI{105}{\hertz}) and simplifies interfacing in large-scale neuromorphic systems. Overall, our neuron offers a unique combination of functional richness, biological plausibility, and energy efficiency, enabled by the choice of a fully analog current-mode implementation of a robust and tunable phase-portrait-based model.

The relatively large area of our neuron ($\sim\SI{3800}{\micro\meter\squared}$) is primarily driven by the capacitors implementing the slow and ultraslow \gls{dpi} filters, which are essential to implement biologically relevant time constants in subthreshold operation and separate the timescales of the mixed-feedback model by a few orders of magnitude. This reflects a fundamental trade-off between biophysical realism and integration density. Several circuit-level techniques could in principle reduce the required capacitor area, including active capacitance multiplication~\cite{razavi_design_2017}, switched-capacitor time constant emulation~\cite{carusone_analog_2012}, and bulk biasing to reduce leakage currents~\cite{qiao_automatic_2019}. These approaches reduce physical capacitance area at the cost of additional circuit complexity, increased power consumption, potential deviations from ideal dynamics, added noise sources, and increased sensitivity to mismatch. The choice between explicit capacitor-based integration and more compact circuit techniques therefore reflects a broader trade-off between density and robustness of the neuron’s dynamical behavior. More radical density improvements may arise from emerging neuromorphic devices such as memristive devices~\cite{Berdan_etal16, moon2024leveraging} now available in \gls{cmos} technology nodes~\cite{atia_hybrid_2025}, and novel uses of existing devices for physics-based integration mechanisms~\cite{pazos_synaptic_2025}. Such approaches could partially replace large explicit capacitors by exploiting intrinsic device dynamics for integration and memory, offering promising directions for future implementations of the mixed-feedback architecture.

Technology scaling is another important consideration that offers both benefits and constraints. While migration to more advanced, smaller nodes would reduce dynamic energy consumption due to lower supply voltages and smaller parasitic capacitances, it limits voltage headroom and typically increases leakage currents~\cite{sood_advanced_2018}. Moreover, in subthreshold analog circuits, transistor dimensions are primarily dictated by mismatch and noise constraints, and capacitor density does not scale proportionally with feature size. Migrating to smaller nodes would therefore not necessarily result in proportional area reduction. As a result, while the mixed-feedback architecture is in principle compatible with advanced nodes, energy and area scaling will depend on careful circuit design optimization, although energy per spike is generally expected to decrease.

\subsection{Interfacing and system-level considerations}\label{subsec:interfacing}
While this work focuses on the design and characterization of a single neuromorphic neuron, practical deployment requires interconnection with other neurons and synapses through a suitable communication scheme. Many neuromorphic systems address this using event-based protocols such as \acrfull{aer} or continuous-time current-mode interconnects in more tightly coupled architectures. While the energy and area overhead associated with these interfaces can be significant at scale, they are largely orthogonal to the intrinsic neuron dynamics presented here and depend strongly on communication sparsity and system-level design trade-offs. The neuron-level energy per spike reported in this work therefore provides a meaningful and generalizable metric for comparing neuronal primitives, with interface optimization more appropriately addressed at the system design stage.

In the proposed architecture, the output of the fast sigmoidal block provides a natural and reliable indicator of spike onset. Thanks to its sharp, saturating transition, this signal reliably switches state during each action potential, independently of slower adaptation or bursting dynamics. As a result, it can be directly interpreted as a spike event without requiring additional thresholding or edge-detection circuitry. It can then be easily copied, rectified, or level-shifted using simple current-mode circuits, enabling straightforward interfacing with event-based communication schemes such as \gls{aer}, as well as with hybrid continuous-time and digital neuromorphic architectures. This separation between the internal continuous-time dynamics and an explicit spike signaling node facilitates modular integration of the neuron into larger networks while preserving its intrinsic low-power operation.

Beyond single-neuron properties, several features of the current-mode mixed-feedback architecture directly support scalable integration. Synaptic and neuronal interactions are naturally expressed as currents, enabling direct summation without additional conversion or arbitration circuitry. Energy per spike remains low and largely independent of absolute current scaling, which is advantageous in large, sparsely active networks. Moreover, the modular separation between neuron dynamics, spike signaling, and communication interfaces allows interconnect architectures to be optimized independently of the neuron. While these characteristics do not, by themselves, constitute a complete neuromorphic system, they provide concrete architectural advantages that motivate the use of the proposed neuron as a building block for large-scale neuromorphic platforms.
\section{Conclusions}\label{sec:concl}

We have demonstrated a fully analog current-mode implementation of a biologically inspired mixed-feedback neuron model that enables robust and tunable neuromodulation in neuromorphic hardware. The proposed circuit uses two types of subthreshold current-mode blocks interconnected in a feedback configuration, and introduces a positive-feedback inactivation mechanism. The resulting neuron can be tuned using mathematical steady-state analyses and exhibits predictable and reliable excitability and neuromodulation in simulation and hardware. A \SI{180}{\nano\meter} \gls{cmos} prototype exhibits spiking, bursting, and real-time neuromodulation between the two using a single bias parameter across wide temperature ranges, validating the model and demonstrating the robustness of the current-mode mixed-feedback approach to global current variations. The neuron operates at \qtyrange{40}{200}{\pico\joule}/spike at a \SI{1.8}{\volt} supply, confirming ultra-low-power operation at the single neuron level and suitability for integration into large-scale neuromorphic systems. Monte Carlo simulations further demonstrate that uniform current scaling provides an effective design knob to trade ultra-low-power operation for substantially improved regime yield and frequency consistency under device mismatch. Overall, the proposed architecture demonstrates that reliable neuromodulable bursting and spiking dynamics can be achieved with current-mode design principles and scalable energy operation. By combining analytical grounding, silicon validation, and robustness evaluation, this work establishes the proposed neuron as a practical and scalable building block for energy-efficient neuromorphic systems.

Future work will showcase the capabilities of networks of mixed-feedback neurons interconnected through synaptic and neuromodulatory interactions to enable adaptive signal processing and sensorimotor control applications. A forthcoming chip will integrate an array of the mixed-feedback neurons presented here, with a revised layout to minimize mismatch, and programmable synaptic connections and neuromodulatory pathways to demonstrate network-level behavioral modulation. This new chip will also enable experimental validation of the energy per spike and power consumption metrics presented in this work. In parallel, we will investigate closed-loop neuromodulation controllers, such as the one presented in~\cite{fyon_reliable_2023}, to autonomously tune bias currents and compensate for device mismatch using knowledge of the neuron's mathematical model. Together, these developments aim to advance mixed-feedback neuromorphic hardware from single-neuron demonstrations towards adaptive, energy-efficient analog computing systems.

\ack{We gratefully acknowledge Philipp Klein for his advice and suggestions on the neuron layout, François Piron for his work on the floorplanning of the test chip and his support during the tapeout phase, and Guillaume Drion for his feedback and advice on the modeling aspects of this work.}

\funding{This work was supported by the Belgian Government through the Federal Public Service Policy and Support grant NEMODEI. Loris Mendolia is a FRIA Grantee of the Fonds de la Recherche Scientifique - FNRS. Giacomo Indiveri was supported by the HORIZON EUROPE EIC Pathﬁnder Grant ELEGANCE (Grant No. 101161114). Elisabetta Chicca would like to acknowledge the financial support of the CogniGron research center and the Ubbo Emmius Funds (Univ. of Groningen).}

\roles{
\begin{itemize}[leftmargin=*]
    \item Loris Mendolia: Conceptualization (lead), Data curation, Formal analysis (lead), Investigation (lead), Methodology (equal), Software, Visualization (lead), Writing - original draft (lead), Writing - review and editing (lead).
    \item Chenxi Wen: Conceptualization (supporting), Investigation (supporting), Methodology (supporting), Writing - review and editing (supporting).
    \item Elisabetta Chicca: Conceptualization (supporting), Investigation (supporting), Resources (supporting), Writing - review and editing (supporting).
    \item Giacomo Indiveri: Conceptualization (supporting), Investigation (supporting), Writing - review and editing (supporting).
    \item Rodolphe Sepulchre: Conceptualization (supporting), Writing - review and editing (supporting).
    \item Jean-Michel Redouté: Conceptualization (supporting), Funding acquisition (equal), Investigation (supporting), Project administration (equal), Resources (equal), Supervision (equal), Writing - original draft (supporting), Writing - review and editing (supporting).
    \item Alessio Franci: Conceptualization (lead), Formal analysis (supporting), Funding acquisition (equal), Investigation (supporting), Methodology (equal), Project administration (equal), Resources (equal), Supervision (equal), Visualization (supporting), Writing - original draft (supporting), Writing - review and editing (lead).
\end{itemize}
}

\data{All data that support the findings of this study are included within the article.}

\newpage
\suppdata{\appendix

\section{Full circuit schematic}\label{circuitappendix}
The full schematic of the current-mode mixed-feedback neuron presented in this work is displayed in \cref{fig:fullschematic}. Transistor dimensions and capacitor sizes used in a revised layout optimized for mismatch are reported in \cref{tab:dimensions}. The dimensions of all transistors of the test circuits used to obtain the experimental results in \cref{sec:results} of this work are $W\times L=\SI{3}{\micro\meter}\times \SI{300}{\nano\meter}$.

\begin{table}[H]
    \centering
    \small
    \begin{tabular}{c|c|c|c|c|c|c|c|c|c|c}
        Transistors & D$_f$1,2 & D$_{s,u}$1,2 & D3 & D$_f$4 & D$_{s,u}$4 & S1 & S2,5,7 & S3,4,6 & M1,2,5 & M3,4\\
        \hline
        $W\times L$ (\si{\micro\meter}) & $1\times 3$ & $2.5\times 4$ & $4\times 5$ & $2\times 2$ & $5\times 5$ & $3\times 5$ & $4\times 5$ & $2\times 0.6$ & $4\times 5$ & $2\times 5$
    \end{tabular}
    
    \begin{tabular}{c|c|c|c}
        Capacitor & $C_f$ & $C_s$ & $C_u$\\
         \hline
        Total capacitance & \SI{0.5}{\pico\farad} & \SI{2}{\pico\farad} & \SI{8}{\pico\farad}
    \end{tabular}
    \caption{Mixed-feedback neuron circuit component dimensions and values (from a mismatch-optimized layout).}
    \label{tab:dimensions}
\end{table}

\begin{figure}[H]
    \centering
    \includegraphics[width=\textwidth]{Figures/fullschematic.tikz}
    \caption{Full current-mode mixed-feedback neuron schematic, implementing the structure detailed in \cref{subsec:modelintro} and in \cref{fig:neuron_diagram}, using the sub-circuits presented in \cref{subsec:circuitblocks} and shown in \cref{fig:DPI_circuit,fig:sigmoid_circuit}. All transistors operate in the subthreshold regime. The fast \gls{dpi} (D$_f$1-4, yellow) models the passive membrane dynamics of the neuron. The fast current-mode sigmoid (S$_f$1-7, orange) provides fast positive feedback on the membrane potential to create the spike upstroke, corresponding to rapid sodium channel dynamics. The slow and ultraslow \glspl{dpi} (D$_{s/u}$1-4, blue and green) introduce linear slow and ultraslow negative feedback to the system, providing repolarization and spike-frequency adaptation currents to the membrane respectively, capturing the effects of delayed-rectifier and calcium-activated potassium currents \cite{hodgkin_quantitative_1952,izhikevich_dynamical_2006}. The slow current-mode sigmoid (S$_s$1-7, pink) provides slow positive feedback on the membrane to create slow regenerative dynamics, mimicking low-threshold calcium currents which support bursting or regenerative depolarization. The positive feedback inactivation circuits (M$_{f/s}$1-5) in the sigmoid blocks subtract copies of the slow and ultraslow currents from the base gain current of the fast and slow sigmoids respectively, inspired by biological sodium channel inactivation observed following spike initiation \cite{zang_sodium_2023}. All unlabeled transistors are part of current mirrors that copy the currents associated with the highlighted gate voltages to implement the various feedback loops of the model.}
    \label{fig:fullschematic}
\end{figure}

\section{Nominal bias parameters}\label{biasesappendix}
The nominal bias parameters used for all simulations of our mixed-feedback neuron presented in this work are reported in \cref{tab:biases}. For experimental results, these currents were converted to bias voltages and applied directly to the corresponding transistor gates.
\begin{table}[H]
    \centering
    \small
    \begin{tabular}{c|c|c|c|c|c|c|c|c}
        $I_{\tau_f}$, $I_{\text{th}_f}$ & $I_{\tau_s}$, $I_{\text{th}_s}$ & $I_{\tau_u}$, $I_{\text{th}_u}$ & $I_{\text{thr}_f}$ & $I_{\text{lin}_f}$ & $I_{\text{G,0}_f}$ & $I_{\text{thr}_s}$ & $I_{\text{lin}_s}$ & $I_{\text{G,0}_s}$ (spk./bst.)\\
        \hline
        \SI{100}{\pico\ampere} & \SI{20}{\pico\ampere} & \SI{10}{\pico\ampere} & \SI{110}{\pico\ampere} & \SI{250}{\pico\ampere} & \SI{700}{\pico\ampere} & \SI{90}{\pico\ampere} & \SI{50}{\pico\ampere} & \SI{200}{\pico\ampere}/\SI{270}{\pico\ampere}
    \end{tabular}
    \caption{Nominal bias currents used in all simulations and approximated in all experiments reported in this work.}
    \label{tab:biases}
\end{table}

\section{Sigmoid circuit equations}\label{sigmoidappendix}

Writing the standard subthreshold MOSFET equations without channel length modulation effect (eq. (3.2.15 of~\cite{liu_analog_2002}) for transistors Q1-7 of \cref{fig:sigmoid_circuit}, only assuming saturation for transistor Q6, yields
\begin{align}
    \text{Q1: }& I_\text{Q1} = I_{0} e^{\kappa\left(V_{DD}-V_\text{in}\right) / U_T}\left(1-e^{-\left(V_{DD}-V_\text{cmp}\right) / U_T}\right)\,,\label{eq1-1}\\
    \text{Q2: }& I_\text{Q2} = I_{0} e^{\kappa V_\text{thr} / U_T}\left(1-e^{-V_\text{cmp} / U_T}\right)\,,\label{eq1-2}\\
    \text{Q3: }& I_\text{cas} = I_{0} e^{\left(\kappa V_\text{cmp} - V_\text{out} )\right/ U_T}\left(1-e^{-\left(V_\text{cmp}-V_\text{out}\right) / U_T}\right)\label{eq1-3}\,,\\
    \text{Q4: }& I_\text{cas} = I_{0} e^{\left(\kappa V_\text{out} - V_\text{cas} \right)/ U_T}\left(1-e^{-\left(V_\text{out}-V_\text{cas}\right) / U_T}\right)\label{eq1-4}\,,\\
    \text{Q5: }& I_\text{cas} = I_{0} e^{\kappa V_\text{lin} / U_T}\left(1-e^{-V_\text{cas} / U_T}\right)\,,\label{eq1-5}\\
    \text{Q6: }& I_\text{out} = I_{0} e^{\left(\kappa V_\text{out} - V_\text{mid} \right)/ U_T}\label{eq1-6}\,,\\
    \text{Q7: }& I_\text{out} = I_{0} e^{\kappa V_\text{G}/ U_T}\left(1-e^{-V_\text{mid} / U_T}\right)\label{eq1-7}\,.
\end{align}

writing the KCL at the common drain between Q1 and Q2 also yields
\begin{equation}
    I_\text{cas} = I_\text{Q1} - I_\text{Q2}\,.\label{kcl}
\end{equation}

The input voltage $V_{in}$ can be assumed to come from a properly saturated PMOS transistor:
\begin{equation}
    I_\text{in} = I_0 e^{\kappa (V_{DD}-V_\text{in})/U_T}\,.\label{eq:siginput}
\end{equation}

From \cref{eq1-6,eq1-7}, the output current can be expressed as a function of $V_\text{out}$ as
\begin{equation}
I_\text{out} = I_0 \frac{e^{\kappa V_\text{G}/U_T}}{1+e^{\kappa\left(V_\text{G} - V_\text{out}\right)/U_T}}\,,\label{eq:sigout}
\end{equation}
which takes the shape of a shifted and scaled logistic function $\mathcal{S}(V_\text{out})=k\frac{1}{1+e^{-V_\text{out}}}+b$. If the uniqueness and monotonicity of a solution for $V_\text{out}$ as a function of $V_\text{in}$ can be proven, this justifies the sigmoidal input-output characteristic of the circuit.

The three devices Q3-5 form a series stack carrying a common current $I_\text{cas}$. At this stage, $I_\text{cas}$ is treated as an unknown scalar parameter.

\begin{itemize}
    \item For a given $I_\text{cas}$, \eqref{eq1-5} uniquely determines $V_\text{cas}$, since the right-hand side is continuous and strictly increasing in $V_\text{cas}$.
    \item With $V_\text{cas}$ fixed, \eqref{eq1-4} uniquely determines $V_\text{out}$, as the current is strictly increasing in $V_\text{out}$.
    \item With $V_\text{out}$ fixed, \eqref{eq1-3} uniquely determines $V_\text{cmp}$, as the current is strictly increasing in $V_\text{cmp}$.
\end{itemize}

Thus, the cascode chain defines unique, strictly increasing functions
\begin{equation}
    V_{cas} = f_5(I_\text{cas})\,, \qquad
    V_{out} = f_4(I_\text{cas})\,, \qquad
    V_{cmp} = f_3(I_\text{cas})\,.
\end{equation}

Substituting \eqref{eq1-1} and \eqref{eq1-2} into \eqref{kcl} yields

\begin{equation}
    I_\text{cas} = I_{0} e^{\kappa\left(V_{DD}-V_\text{in}\right) / U_T}\left(1-e^{-\left(V_{DD}-V_\text{cmp}\right) / U_T}\right) - I_{0} e^{\kappa V_\text{thr} / U_T}\left(1-e^{-V_\text{cmp} / U_T}\right)\,.\label{eq:Icas_app}
\end{equation}

The right-hand side of \eqref{eq:Icas_app} is continuous and strictly increasing in $V_\text{cmp}$, and strictly decreasing in $V_\text{in}$.

Substituting $V_\text{cmp}=f_3(I_\text{cas})$ into \eqref{eq:Icas_app} yields a single implicit equation in \(I_{cas}\):
\begin{equation}
   I_\text{cas} = I_{0} e^{\kappa\left(V_{DD}-V_\text{in}\right) / U_T}\left(1-e^{-\left(V_{DD}-f_3(I_\text{cas})\right) / U_T}\right) - I_{0} e^{\kappa V_\text{thr} / U_T}\left(1-e^{-f_3(I_\text{cas}) / U_T}\right)\,.
\end{equation}
This equation does not admit a closed-form solution in elementary functions.

For fixed $V_\text{in}$, both sides of the implicit equation are continuous and strictly increasing functions of $I_\text{cas}$. Therefore, a unique solution for $I_\text{cas}$ exists. Since $V_{out} = f_4(I_\text{cas})$, the output voltage is also uniquely defined.

Moreover, because $I_\text{cas}$ is strictly decreasing in $V_\text{in}$ and $V_\text{out}$ is strictly increasing in $I_\text{cas}$, the input-output characteristic satisfies
\begin{equation}
    \frac{dV_\text{out}}{dV_\text{in}} < 0,
\end{equation}
\textit{i.e.}, the input-output characteristic is strictly monotonic.

Adding that $V_\text{in}$ is strictly increasing in $I_\text{in}$ from \eqref{eq:siginput}, and with at the $V_\text{out}$ to $I_\text{out}$ relationship in \eqref{eq:sigout}, this proves that our sigmoid circuit implements a strictly monotonic and increasing scaled and shifted logistic sigmoidal relationship between the input current $I_\text{in}$ and the output current $I_\text{out}$. \qed}

\bibliographystyle{ieeetr}
\bibliography{references}

@article{franci_robust_2018,
	title = {Robust and tunable bursting requires slow positive feedback},
	volume = {119},
	issn = {0022-3077, 1522-1598},
	doi = {10.1152/jn.00804.2017},
	language = {en},
	number = {3},
	journal = {Journal of Neurophysiology},
	author = {Franci, Alessio and Drion, Guillaume and Sepulchre, Rodolphe},
	month = mar,
	year = {2018},
	pages = {1222--1234},
}

@article{drion_dynamic_2015,
	title = {Dynamic {Input} {Conductances} {Shape} {Neuronal} {Spiking}},
	volume = {2},
	issn = {2373-2822},
	doi = {10.1523/ENEURO.0031-14.2015},
	language = {en},
	number = {1},
	journal = {eneuro},
	author = {Drion, Guillaume and Franci, Alessio and Dethier, Julie and Sepulchre, Rodolphe},
	month = jan,
	year = {2015},
	pages = {ENEURO.0031--14.2015},
}

@inproceedings{livi_current-mode_2009,
	title = {A current-mode conductance-based silicon neuron for address-event neuromorphic systems},
	doi = {10.1109/ISCAS.2009.5118408},
	booktitle = {2009 {IEEE} {International} {Symposium} on {Circuits} and {Systems}},
	author = {Livi, Paolo and Indiveri, Giacomo},
	month = may,
	year = {2009},
	note = {ISSN: 2158-1525},
	pages = {2898--2901},
}

@article{marder_neuromodulation_2012,
	title = {Neuromodulation of {Neuronal} {Circuits}: {Back} to the {Future}},
	volume = {76},
	issn = {0896-6273},
	shorttitle = {Neuromodulation of {Neuronal} {Circuits}},
	doi = {10.1016/j.neuron.2012.09.010},
	language = {en},
	number = {1},
	journal = {Neuron},
	author = {Marder, Eve},
	month = oct,
	year = {2012},
	pages = {1--11},
}

@article{ribar_neuromodulation_2019,
	title = {Neuromodulation of {Neuromorphic} {Circuits}},
	volume = {66},
	issn = {1549-8328, 1558-0806},
	doi = {10.1109/TCSI.2019.2907113},
	number = {8},
	journal = {{IEEE} Transactions on Circuits and Systems---Part {I}: Regular Papers},
	author = {Ribar, Luka and Sepulchre, Rodolphe},
	month = aug,
	year = {2019},
	pages = {3028--3040},
}

@article{sherman_tonic_2001,
    title = {Tonic and burst firing: dual modes of thalamocortical relay},
    volume = {24},
    copyright = {https://www.elsevier.com/tdm/userlicense/1.0/},
    issn = {01662236},
    shorttitle = {Tonic and burst firing},
    doi = {10.1016/S0166-2236(00)01714-8},
    language = {en},
    number = {2},
    journal = {Trends in Neurosciences},
    author = {Sherman, S.Murray},
    month = feb,
    year = {2001},
    pages = {122--126},
}

@article{zang_sodium_2023,
    title = {Sodium channel slow inactivation normalizes firing in axons with uneven conductance distributions},
    volume = {33},
    issn = {0960-9822},
    doi = {10.1016/j.cub.2023.03.043},
    number = {9},
    journal = {Current Biology},
    author = {Zang, Yunliang and Marder, Eve and Marom, Shimon},
    month = may,
    year = {2023},
    pages = {1818--1824.e3},
}

@article{bargmann_beyond_2012,
    title = {Beyond the connectome: how neuromodulators shape neural circuits},
    volume = {34},
    issn = {1521-1878},
    shorttitle = {Beyond the connectome},
    doi = {10.1002/bies.201100185},
    language = {eng},
    number = {6},
    journal = {BioEssays: News and Reviews in Molecular, Cellular and Developmental Biology},
    author = {Bargmann, Cornelia I.},
    month = jun,
    year = {2012},
    pmid = {22396302},
    pages = {458--465},
}

@article{liu_bionn_2023,
    title = {{BioNN}: {Bio}-{Mimetic} {Neural} {Networks} on {Hardware} {Using} {Nonlinear} {Multi}-{Timescale} {Mixed}-{Feedback} {Control} for {Neuromodulatory} {Bursting} {Rhythms}},
    volume = {13},
    issn = {2156-3357, 2156-3365},
    shorttitle = {{BioNN}},
    doi = {10.1109/JETCAS.2023.3330084},
    language = {en},
    number = {4},
    journal = {{IEEE} Journal on Emerging and Selected Topics in Circuits and Systems},
    author = {Liu, Kangni and Hashemkhani, Shahin and Rubin, Jonathan and Kubendran, Rajkumar},
    month = dec,
    year = {2023},
    pages = {914--926},
}

@article{mead_neuromorphic_1990,
    title = {Neuromorphic electronic systems},
    volume = {78},
    copyright = {https://ieeexplore.ieee.org/Xplorehelp/downloads/license-information/IEEE.html},
    issn = {00189219},
    doi = {10.1109/5.58356},
    number = {10},
    journal = {Proceedings of the {IEEE}},
    author = {Mead, C.},
    month = oct,
    year = {1990},
    pages = {1629--1636},
}

@article{hodgkin_quantitative_1952,
    title = {A quantitative description of membrane current and its application to conduction and excitation in nerve},
    volume = {117},
    issn = {0022-3751},
    doi = {10.1113/jphysiol.1952.sp004764},
    number = {4},
    journal = {The Journal of Physiology},
    author = {Hodgkin, A. L. and Huxley, A. F.},
    month = aug,
    year = {1952},
    pmid = {12991237},
    pmcid = {PMC1392413},
    pages = {500--544},
}

@book{izhikevich_dynamical_2006,
    title = {Dynamical {Systems} in {Neuroscience}: {The} {Geometry} of {Excitability} and {Bursting}},
    isbn = {978-0-262-27607-8},
    shorttitle = {Dynamical {Systems} in {Neuroscience}},
    language = {en},
    publisher = {The MIT Press},
    author = {Izhikevich, Eugene M.},
    month = jul,
    year = {2006},
    doi = {10.7551/mitpress/2526.001.0001},
}

@article{wijekoon_compact_2008,
    series = {Advances in {Neural} {Networks} {Research}: {IJCNN} ’07},
    title = {Compact silicon neuron circuit with spiking and bursting behaviour},
    volume = {21},
    issn = {0893-6080},
    doi = {10.1016/j.neunet.2007.12.037},
    number = {2},
    journal = {Neural Networks},
    author = {Wijekoon, Jayawan H. B. and Dudek, Piotr},
    month = mar,
    year = {2008},
    pages = {524--534},
}

@inproceedings{hynna_silicon_2007,
    title = {Silicon neurons that burst when primed},
    doi = {10.1109/ISCAS.2007.378288},
    booktitle = {2007 {IEEE} {International} {Symposium} on {Circuits} and {Systems} ({ISCAS})},
    author = {Hynna, Kai M and Boahen, Kwabena},
    month = may,
    year = {2007},
    note = {ISSN: 2158-1525},
    pages = {3363--3366},
}

@inproceedings{wang_compact_2015,
    address = {Atlanta, GA, USA},
    title = {A compact {aVLSI} conductance-based silicon neuron},
    doi = {10.1109/biocas.2015.7348396},
    language = {en},
    booktitle = {2015 {IEEE} {Biomedical} {Circuits} and {Systems} {Conference} ({BioCAS})},
    publisher = {IEEE},
    author = {Wang, Runchun and Thakur, Chetan Singh and Hamilton, Tara Julia and Tapson, Jonathan and Van Schaik, Andre},
    month = oct,
    year = {2015},
    pages = {1--4},
}

@misc{pehle_brainscales-2_2022,
    title = {The {BrainScaleS}-2 accelerated neuromorphic system with hybrid plasticity},
    doi = {10.48550/arXiv.2201.11063},
    urldate = {2025-07-17},
    publisher = {arXiv},
    author = {Pehle, Christian and Billaudelle, Sebastian and Cramer, Benjamin and Kaiser, Jakob and Schreiber, Korbinian and Stradmann, Yannik and Weis, Johannes and Leibfried, Aron and Müller, Eric and Schemmel, Johannes},
    month = feb,
    year = {2022},
    note = {arXiv:2201.11063 [cs]},
}

@article{liu_artificial_2023,
    title = {Artificial {Spiking} {Neuron} with {Bursting} {Dynamics} for {Noise}-{Resistant} {Neuromorphic} {Coding}},
    volume = {5},
    copyright = {https://doi.org/10.15223/policy-029},
    issn = {2637-6113, 2637-6113},
    doi = {10.1021/acsaelm.3c00445},
    language = {en},
    number = {6},
    journal = {ACS Applied Electronic Materials},
    author = {Liu, Huiyuan and Zhu, Xiaojian and Guo, Zhecheng and He, Ri and Li, Xinze and Sun, Qihao and Ye, Xiaoyu and Sun, Cui and Tian, Yu and Li, Run-Wei},
    month = jun,
    year = {2023},
    publisher = {American Chemical Society (ACS)},
    pages = {3454--3461},
}

@article{xiao_bio-plausible_2025,
    title = {Bio-plausible reconfigurable spiking neuron for neuromorphic computing},
    volume = {11},
    doi = {10.1126/sciadv.adr6733},
    number = {6},
    journal = {Science Advances},
    author = {Xiao, Yu and Liu, Yize and Zhang, Bihua and Chen, Peng and Zhu, Huaze and He, Enhui and Zhao, Jiayi and Huo, Wenju and Jin, Xiaofei and Zhang, Xumeng and Jiang, Hao and Ma, De and Zheng, Qian and Tang, Huajin and Lin, Peng and Kong, Wei and Pan, Gang},
    month = feb,
    year = {2025},
    pages = {eadr6733},
}

@article{folowosele_silicon_2011,
    title = {Silicon {Modeling} of the {Mihalaş}–{Niebur} {Neuron}},
    volume = {22},
    issn = {1941-0093},
    doi = {10.1109/TNN.2011.2167020},
    number = {12},
    urldate = {2025-07-18},
    journal = {{IEEE} Transactions on Neural Networks},
    author = {Folowosele, Fopefolu and Hamilton, Tara Julia and Etienne-Cummings, Ralph},
    month = dec,
    year = {2011},
    pages = {1915--1927},
}

@article{yu_biophysical_2011,
    title = {Biophysical {Neural} {Spiking}, {Bursting}, and {Excitability} {Dynamics} in {Reconfigurable} {Analog} {VLSI}},
    volume = {5},
    issn = {1940-9990},
    doi = {10.1109/TBCAS.2011.2169794},
    number = {5},
    urldate = {2025-07-24},
    journal = {{IEEE} Transactions on Biomedical Circuits and Systems},
    author = {Yu, Theodore and Sejnowski, Terrence J. and Cauwenberghs, Gert},
    month = oct,
    year = {2011},
    pages = {420--429},
}

@article{aamir_accelerated_2018,
    title = {An {Accelerated} {LIF} {Neuronal} {Network} {Array} for a {Large}-{Scale} {Mixed}-{Signal} {Neuromorphic} {Architecture}},
    volume = {65},
    issn = {1558-0806},
    doi = {10.1109/TCSI.2018.2840718},
    number = {12},
    urldate = {2025-07-24},
    journal = {{IEEE} Transactions on Circuits and Systems---Part {I}: Regular Papers},
    author = {Aamir, Syed Ahmed and Stradmann, Yannik and Müller, Paul and Pehle, Christian and Hartel, Andreas and Grübl, Andreas and Schemmel, Johannes and Meier, Karlheinz},
    month = dec,
    year = {2018},
    pages = {4299--4312},
}

@article{jacquerie_robust_2021,
    title = {Robust switches in thalamic network activity require a timescale separation between sodium and {T}-type calcium channel activations},
    volume = {17},
    issn = {1553-7358},
    doi = {10.1371/journal.pcbi.1008997},
    language = {en},
    number = {5},
    journal = {PLOS Computational Biology},
    author = {Jacquerie, Kathleen and Drion, Guillaume},
    year = {2021},
    publisher = {Public Library of Science},
    pages = {e1008997},
}

@article{drion_novel_2012,
    title = {A {Novel} {Phase} {Portrait} for {Neuronal} {Excitability}},
    volume = {7},
    issn = {1932-6203},
    doi = {10.1371/journal.pone.0041806},
    language = {en},
    number = {8},
    journal = {PLOS ONE},
    author = {Drion, Guillaume and Franci, Alessio and Seutin, Vincent and Sepulchre, Rodolphe},
    year = {2012},
    publisher = {Public Library of Science},
    pages = {e41806},
}

@article{fyon_reliable_2023,
    series = {22nd {IFAC} {World} {Congress}},
    title = {Reliable neuromodulation from adaptive control of ion channel expression},
    volume = {56},
    issn = {2405-8963},
    doi = {10.1016/j.ifacol.2023.10.1610},
    number = {2},
    journal = {IFAC-PapersOnLine},
    author = {Fyon, A. and Sacré, P. and Franci, A. and Drion, G.},
    month = jan,
    year = {2023},
    pages = {458--463},
}

@article{chicca_neuromorphic_2014,
    title = {Neuromorphic {Electronic} {Circuits} for {Building} {Autonomous} {Cognitive} {Systems}},
    volume = {102},
    issn = {0018-9219, 1558-2256},
    doi = {10.1109/JPROC.2014.2313954},
    language = {en},
    number = {9},
    urldate = {2022-10-27},
    journal = {Proceedings of the {IEEE}},
    author = {Chicca, Elisabetta and Stefanini, Fabio and Bartolozzi, Chiara and Indiveri, Giacomo},
    month = sep,
    year = {2014},
    pages = {1367--1388},
}

@article{avery_neuromodulatory_2017,
    title = {Neuromodulatory {Systems} and {Their} {Interactions}},
    volume = {11},
    issn = {1662-5110},
    shorttitle = {Neuromodulatory {Systems} and {Their} {Interactions}},
    doi = {10.3389/fncir.2017.00108},
    journal = {Frontiers in Neural Circuits},
    author = {Avery, Michael C. and Krichmar, Jeffrey L.},
    month = dec,
    year = {2017},
    pmid = {29311844},
    pmcid = {PMC5744617},
    pages = {108},
}

@article{mccormick_editorial_2014,
    title = {Editorial overview: {Neuromodulation}: {Tuning} the properties of neurons, networks and behavior},
    volume = {29},
    issn = {0959-4388},
    shorttitle = {Editorial overview},
    doi = {10.1016/j.conb.2014.10.010},
    journal = {Current opinion in neurobiology},
    author = {McCormick, David A and Nusbaum, Michael P},
    month = dec,
    year = {2014},
    pmid = {25457725},
    pmcid = {PMC4450677},
    pages = {iv--vii},
}

@misc{mei_improving_2025,
    title = {Improving the adaptive and continuous learning capabilities of artificial neural networks: {Lessons} from multi-neuromodulatory dynamics},
    shorttitle = {Improving the adaptive and continuous learning capabilities of artificial neural networks},
    doi = {10.48550/arXiv.2501.06762},
    publisher = {arXiv},
    author = {Mei, Jie and Rodriguez-Garcia, Alejandro and Takeuchi, Daigo and Wainstein, Gabriel and Hubig, Nina and Mohsenzadeh, Yalda and Ramaswamy, Srikanth},
    month = jan,
    year = {2025},
    note = {arXiv:2501.06762 [q-bio]},
}

@article{yang_neuromorphic_2020,
    title = {Neuromorphic {Engineering}: {From} {Biological} to {Spike}-{Based} {Hardware} {Nervous} {Systems}},
    volume = {32},
    copyright = {© 2020 Wiley-VCH GmbH},
    issn = {1521-4095},
    shorttitle = {Neuromorphic {Engineering}},
    doi = {10.1002/adma.202003610},
    language = {en},
    number = {52},
    journal = {Advanced Materials},
    author = {Yang, Jia-Qin and Wang, Ruopeng and Ren, Yi and Mao, Jing-Yu and Wang, Zhan-Peng and Zhou, Ye and Han, Su-Ting},
    year = {2020},
    pages = {2003610},
}

@misc{schuman_survey_2017,
    title = {A {Survey} of {Neuromorphic} {Computing} and {Neural} {Networks} in {Hardware}},
    doi = {10.48550/arXiv.1705.06963},
    publisher = {arXiv},
    author = {Schuman, Catherine D. and Potok, Thomas E. and Patton, Robert M. and Birdwell, J. Douglas and Dean, Mark E. and Rose, Garrett S. and Plank, James S.},
    month = may,
    year = {2017},
    note = {arXiv:1705.06963 [cs]},
}

@article{poon_neuromorphic_2011,
    title = {Neuromorphic {Silicon} {Neurons} and {Large}-{Scale} {Neural} {Networks}},
    volume = {5},
    issn = {1662-4548},
    shorttitle = {Neuromorphic {Silicon} {Neurons} and {Large}-{Scale} {Neural} {Networks}},
    doi = {10.3389/fnins.2011.00108},
    urldate = {2025-08-07},
    journal = {Frontiers in Neuroscience},
    author = {Poon, Chi-Sang and Zhou, Kuan},
    month = sep,
    year = {2011},
    pmid = {21991244},
    pmcid = {PMC3181466},
    pages = {108},
}

@article{neftci_data_2018,
    title = {Data and {Power} {Efficient} {Intelligence} with {Neuromorphic} {Learning} {Machines}},
    volume = {5},
    issn = {2589-0042},
    doi = {10.1016/j.isci.2018.06.010},
    journal = {iScience},
    author = {Neftci, Emre O.},
    month = jul,
    year = {2018},
    pages = {52--68},
}

@article{ijspeert_central_2008,
    series = {Robotics and {Neuroscience}},
    title = {Central pattern generators for locomotion control in animals and robots},
    volume = {21},
    issn = {0893-6080},
    shorttitle = {Central pattern generators for locomotion control in animals and robots},
    doi = {10.1016/j.neunet.2008.03.014},
    language = {en},
    number = {4},
    journal = {Neural Networks},
    author = {Ijspeert, Auke Jan},
    month = may,
    year = {2008},
    pages = {642--653},
}

@article{stuck_burst-dependent_2025,
    title = {A burst-dependent algorithm for neuromorphic on-chip learning of spiking neural networks},
    volume = {5},
    issn = {2634-4386},
    doi = {10.1088/2634-4386/adb511},
    language = {en},
    number = {1},
    urldate = {2025-07-28},
    journal = {Neuromorphic Computing and Engineering},
    author = {Stuck, Michael and Wang, Xingyun and Naud, Richard},
    month = feb,
    year = {2025},
    publisher = {IOP Publishing},
    pages = {014010},
}

@article{li_bursting_2016,
    title = {Bursting dynamics remarkably improve the performance of neural networks on liquid computing},
    volume = {10},
    issn = {1871-4080},
    doi = {10.1007/s11571-016-9387-z},
    number = {5},
    urldate = {2025-08-07},
    journal = {Cognitive Neurodynamics},
    author = {Li, Xiumin and Chen, Qing and Xue, Fangzheng},
    month = oct,
    year = {2016},
    pmid = {27668020},
    pmcid = {PMC5018008},
    pages = {415--421},
}

@misc{donati_neuromorphic_2021,
    title = {Neuromorphic {Pattern} {Generation} {Circuits} for {Bioelectronic} {Medicine}},
    doi = {10.48550/arXiv.2102.09630},
    publisher = {arXiv},
    author = {Donati, Elisa and Krause, Renate and Indiveri, Giacomo},
    month = feb,
    year = {2021},
    note = {arXiv:2102.09630 [cs]},
}

@inproceedings{athota_neuromorphic_2021,
    title = {Neuromorphic {Instantiation} of {Spiking} {Half}-{Centered} {Oscillator} {Models} for {Central} {Pattern} {Generation}},
    isbn = {978-1-72811-179-7},
    doi = {10.1109/EMBC46164.2021.9629606},
    language = {en},
    urldate = {2024-01-30},
    booktitle = {2021 43rd {Annual} {International} {Conference} of the {IEEE} {Engineering} in {Medicine} \& {Biology} {Society} ({EMBC})},
    author = {Athota, Aditya and Caccam, Blair and Kochis, Ryan and Ray, Arjun and Cauwenberghs, Gert and Broccard, Frederic D.},
    month = nov,
    year = {2021},
    pages = {6703--6706},
}

@article{lopez-osorio_neuromorphic_2022,
    title = {Neuromorphic adaptive spiking {CPG} towards bio-inspired locomotion},
    volume = {502},
    issn = {0925-2312},
    doi = {10.1016/j.neucom.2022.06.085},
    language = {en},
    urldate = {2024-01-29},
    journal = {Neurocomputing},
    author = {Lopez-Osorio, Pablo and Patiño-Saucedo, Alberto and Dominguez-Morales, Juan P. and Rostro-Gonzalez, Horacio and Perez-Peña, Fernando},
    month = sep,
    year = {2022},
    pages = {57--70},
}

@article{bartolozzi_embodied_2022,
    title = {Embodied neuromorphic intelligence},
    volume = {13},
    issn = {2041-1723},
    doi = {10.1038/s41467-022-28487-2},
    language = {en},
    number = {1},
    urldate = {2022-10-27},
    journal = {Nature Communications},
    author = {Bartolozzi, Chiara and Indiveri, Giacomo and Donati, Elisa},
    month = feb,
    year = {2022},
    pages = {1024},
}

@article{deweerth_simple_1991,
    title = {A simple neuron servo},
    volume = {2},
    issn = {1941-0093},
    doi = {10.1109/72.80335},
    number = {2},
    journal = {{IEEE} Transactions on Neural Networks},
    author = {DeWeerth, S.P. and Nielsen, L. and Mead, C.A. and Astrom, K.J.},
    month = mar,
    year = {1991},
    pages = {248--251},
}

@article{richter_dynap-se2_2024,
    title = {{DYNAP}-{SE2}: a scalable multi-core dynamic neuromorphic asynchronous spiking neural network processor},
    volume = {4},
    issn = {2634-4386},
    shorttitle = {{DYNAP}-{SE2}},
    url = {https://dx.doi.org/10.1088/2634-4386/ad1cd7},
    doi = {10.1088/2634-4386/ad1cd7},
    language = {en},
    number = {1},
    urldate = {2024-04-02},
    journal = {Neuromorphic Computing and Engineering},
    author = {Richter, Ole and Wu, Chenxi and Whatley, Adrian M and Köstinger, German and Nielsen, Carsten and Qiao, Ning and Indiveri, Giacomo},
    month = jan,
    year = {2024},
    note = {Publisher: IOP Publishing},
    pages = {014003},
}

@article{thakur_large-scale_2018,
    title = {Large-{Scale} {Neuromorphic} {Spiking} {Array} {Processors}: {A} {Quest} to {Mimic} the {Brain}},
    volume = {12},
    issn = {1662-4548},
    shorttitle = {Large-{Scale} {Neuromorphic} {Spiking} {Array} {Processors}},
    url = {https://www.ncbi.nlm.nih.gov/pmc/articles/PMC6287454/},
    doi = {10.3389/fnins.2018.00891},
    urldate = {2025-08-13},
    journal = {Frontiers in Neuroscience},
    author = {Thakur, Chetan Singh and Molin, Jamal Lottier and Cauwenberghs, Gert and Indiveri, Giacomo and Kumar, Kundan and Qiao, Ning and Schemmel, Johannes and Wang, Runchun and Chicca, Elisabetta and Olson Hasler, Jennifer and Seo, Jae-sun and Yu, Shimeng and Cao, Yu and van Schaik, André and Etienne-Cummings, Ralph},
    month = dec,
    year = {2018},
    pmid = {30559644},
    pmcid = {PMC6287454},
    pages = {891},
}

@article{drion2019celullar,
	title = {Cellular Switches Orchestrate Rhythmic Circuits},
	author = {Drion, Guillaume and Franci, Alessio and Sepulchre, Rodolphe},
	year = {2019},
	month = apr,
	journal = {Biological Cybernetics},
	volume = {113},
	number = {1-2},
	pages = {71--82},
	issn = {0340-1200, 1432-0770},
	doi = {10.1007/s00422-018-0778-6},
	urldate = {2025-01-01},
	langid = {english}
}

@article{sepulchre2019control,
	title = {Control {{Across Scales}} by {{Positive}} and {{Negative Feedback}}},
	author = {Sepulchre, R. and Drion, G. and Franci, A.},
	year = {2019},
	month = may,
	journal = {Annual Review of Control, Robotics, and Autonomous Systems},
	volume = {2},
	number = {1},
	pages = {89--113},
	issn = {2573-5144, 2573-5144},
	doi = {10.1146/annurev-control-053018-023708},
	urldate = {2025-01-01},
	langid = {english}
}

@article{rubino2021ultralow,
	title = {Ultra-{{Low-Power FDSOI Neural Circuits}} for {{Extreme-Edge Neuromorphic Intelligence}}},
	author = {Rubino, Arianna and Livanelioglu, Can and Qiao, Ning and Payvand, Melika and Indiveri, Giacomo},
	year = {2021},
	month = jan,
	journal = {IEEE Transactions on Circuits and Systems I: Regular Papers},
	volume = {68},
	number = {1},
	pages = {45--56},
	issn = {1558-0806},
	doi = {10.1109/TCSI.2020.3035575},
	urldate = {2025-08-26}
}

@phdthesis{rubino2025Phd,
	title = {Mixed-{{Signal Neuromorphic Circuits}} and {{Systems}} for {{Extreme-Edge Computing}}},
	author = {Rubino, Arianna},
	year = {2025},
	month = mar,
	journal = {Rubino, Arianna. Mixed-Signal Neuromorphic Circuits and Systems for Extreme-Edge Computing.  2025, University of Zurich, Faculty of Science.},
	address = {Z{\"u}rich},
	doi = {10.5167/uzh-276596},
	urldate = {2025-08-26},
	copyright = {info:eu-repo/semantics/closedAccess},
	langid = {english},
	school = {University of Zurich}
}

@article{franci2013balance,
	title = {A {{Balance Equation Determines}} a {{Switch}} in {{Neuronal Excitability}}},
	author = {Franci, Alessio and Drion, Guillaume and Seutin, Vincent and Sepulchre, Rodolphe},
	year = {2013},
	month = may,
	journal = {PLOS Computational Biology},
	volume = {9},
	number = {5},
	pages = {e1003040},
	publisher = {Public Library of Science},
	issn = {1553-7358},
	doi = {10.1371/journal.pcbi.1003040},
	urldate = {2025-08-27},
	langid = {english}
}

@article{castanos2017implementing,
	title = {Implementing Robust Neuromodulation in Neuromorphic Circuits},
	author = {Casta{\~n}os, Fernando and Franci, Alessio},
	year = {2017},
	month = apr,
	journal = {Neurocomputing},
	series = {{{SI}}: {{CCE}} 2015},
	volume = {233},
	pages = {3--13},
	issn = {0925-2312},
	doi = {10.1016/j.neucom.2016.08.099},
	urldate = {2025-08-27}
}

@book{Hille2001,
	author = {B. Hille},
	publisher = {Sinauer Associates Inc.,U.S.},
	title = {Ion Channels of Excitable Membranes},
	year = {2001}}

@inproceedings{bartolozzi_ultra_2006,
    title = {An ultra low power current-mode filter for neuromorphic systems and biomedical signal processing},
    url = {https://ieeexplore.ieee.org/document/4600325},
    doi = {10.1109/BIOCAS.2006.4600325},
    urldate = {2025-08-27},
    booktitle = {2006 {IEEE} {Biomedical} {Circuits} and {Systems} {Conference}},
    author = {Bartolozzi, Chiara and Mitra, Srinjoy and Indiveri, Giacomo},
    month = nov,
    year = {2006},
    note = {ISSN: 2163-4025},
    pages = {130--133},
}

@article{tang_robustness_2012,
    title = {Robustness of a {Rhythmic} {Circuit} to {Short}- and {Long}-{Term} {Temperature} {Changes}},
    volume = {32},
    issn = {0270-6474, 1529-2401},
    url = {https://www.jneurosci.org/lookup/doi/10.1523/JNEUROSCI.1443-12.2012},
    doi = {10.1523/JNEUROSCI.1443-12.2012},
    language = {en},
    number = {29},
    urldate = {2025-08-27},
    journal = {Journal of Neuroscience},
    author = {Tang, L. S. and Taylor, A. L. and Rinberg, A. and Marder, E.},
    month = jul,
    year = {2012},
    pages = {10075--10085},
}

@article{ratliff_neuronal_2021,
    title = {Neuronal oscillator robustness to multiple global perturbations},
    volume = {120},
    issn = {0006-3495, 1542-0086},
    url = {https://www.cell.com/biophysj/abstract/S0006-3495(21)00147-8},
    doi = {10.1016/j.bpj.2021.01.038},
    language = {English},
    number = {8},
    urldate = {2025-08-27},
    journal = {Biophysical Journal},
    author = {Ratliff, Jacob and Franci, Alessio and Marder, Eve and O’Leary, Timothy},
    month = apr,
    year = {2021},
    pmid = {33610580},
    note = {Publisher: Elsevier},
    pages = {1454--1468},
}

@article{bartolozzi_synaptic_2007,
    title = {Synaptic dynamics in analog {VLSI}},
    volume = {19},
    issn = {0899-7667},
    doi = {10.1162/neco.2007.19.10.2581},
    language = {eng},
    number = {10},
    journal = {Neural Computation},
    author = {Bartolozzi, Chiara and Indiveri, Giacomo},
    month = oct,
    year = {2007},
    pmid = {17716003},
    pages = {2581--2603},
}

@Article{Indiveri25,
author		= {Giacomo Indiveri},
title		= {Neuromorphic is dead. Long live neuromorphic},
journal		= {Neuron},
year		= {2025},
month		= oct,
pages		= {1--4},
doi		= {10.1016/j.neuron.2025.09.020}
}

@Article{Wunderlich_etal19,
author		= {Wunderlich, Timo and Kungl, Akos F. and M{\"u}ller, Eric
		  and Hartel, Andreas and Stradmann, Yannik and Aamir, Syed
		  Ahmed and Gr{\"u}bl, Andreas and Heimbrecht, Arthur and
		  Schreiber, Korbinian and St{\"o}ckel, David and Pehle,
		  Christian and Billaudelle, Sebastian and Kiene, Gerd and
		  Mauch, Christian and Schemmel, Johannes and Meier,
		  Karlheinz and Petrovici, Mihai A.},
title		= {Demonstrating Advantages of Neuromorphic Computation: A
		  Pilot Study},
journal		= {Frontiers in Neuroscience},
year		= {2019},
volume		= {13},
issn		= {1662-453X},
doi		= {10.3389/fnins.2019.00260},
url		= {https://www.frontiersin.org/articles/10.3389/fnins.2019.00260}
}

@Article{Berdan_etal16,
author		= {R. Berdan and E. Vasilaki and A. Khiat and G. Indiveri and
		  A. Serb and T. Prodromakis},
title		= {Emulating short-term synaptic dynamics with memristive
		  devices},
journal		= {Scientific {R}eports},
year		= {2016},
volume		= {6},
number		= {18639},
pages		= {1--9},
doi		= {10.1038/srep18639}
}

@article{moon2024leveraging,
  title={Leveraging volatile memristors in neuromorphic computing: from materials to system implementation},
  author={Moon, Taehwan and Soh, Keunho and Kim, Jong Sung and Kim, Ji Eun and Chun, Suk Yeop and Cho, Kyungjune and Yang, J Joshua and Yoon, Jung Ho},
  journal={Materials Horizons},
  volume={11},
  number={20},
  pages={4840--4866},
  year={2024},
  publisher={Royal Society of Chemistry}
}

@article{yu_analog_2010,
    title = {Analog {VLSI} {Biophysical} {Neurons} and {Synapses} {With} {Programmable} {Membrane} {Channel} {Kinetics}},
    volume = {4},
    copyright = {https://ieeexplore.ieee.org/Xplorehelp/downloads/license-information/IEEE.html},
    issn = {1932-4545, 1940-9990},
    url = {http://ieeexplore.ieee.org/document/5471736/},
    doi = {10.1109/TBCAS.2010.2048566},
    number = {3},
    urldate = {2026-01-08},
    journal = {IEEE Transactions on Biomedical Circuits and Systems},
    author = {Yu, Theodore and Cauwenberghs, Gert},
    month = jun,
    year = {2010},
    pages = {139--148},
}

@article{benjamin_analytical_2023,
    title = {An {Analytical} {MOS} {Device} {Model} {With} {Mismatch} and {Temperature} {Variation} for {Subthreshold} {Circuits}},
    volume = {70},
    issn = {1558-3791},
    url = {https://ieeexplore.ieee.org/document/10005201},
    doi = {10.1109/TCSII.2023.3234009},
    number = {6},
    urldate = {2026-02-07},
    journal = {IEEE Transactions on Circuits and Systems II: Express Briefs},
    author = {Benjamin, Ben Varkey and Smith, Richelle L. and Boahen, Kwabena A.},
    month = jun,
    year = {2023},
    pages = {1826--1830},
}

@book{razavi_design_2017,
    address = {New York, NY},
    edition = {Second edition},
    title = {Design of analog {CMOS} integrated circuits},
    isbn = {978-0-07-252493-2 978-981-4636-26-1},
    language = {en},
    publisher = {McGraw-Hill Education},
    author = {Razavi, Behzad},
    year = {2017},
}

@article{sood_advanced_2018,
    title = {Advanced {MOSFET} {Technologies} for {Next} {Generation} {Communication} {Systems} - {Perspective} and {Challenges}: {A} {Review}},
    volume = {11},
    issn = {17919320, 17912377},
    shorttitle = {Advanced {MOSFET} {Technologies} for {Next} {Generation} {Communication} {Systems} - {Perspective} and {Challenges}},
    url = {http://jestr.org/downloads/Volume11Issue3/fulltext251132018.pdf},
    doi = {10.25103/jestr.113.25},
    number = {3},
    urldate = {2026-02-16},
    journal = {Journal of Engineering Science and Technology Review},
    author = {Sood, Himangi and Srivastava, Viranjay M. and Singh, Ghanshyam},
    month = apr,
    year = {2018},
    pages = {180--195},
}

@book{carusone_analog_2012,
    address = {Hoboken, NJ},
    edition = {2nd ed},
    title = {Analog integrated circuit design},
    isbn = {978-0-470-77010-8},
    language = {en},
    publisher = {J. Wiley \& Sons},
    author = {Carusone, Tony Chan},
    collaborator = {Johns, David and Martin, Kenneth W.},
    year = {2012},
}

@misc{qiao_automatic_2019,
    title = {Automatic gain control of ultra-low leakage synaptic scaling homeostatic plasticity circuits},
    url = {http://arxiv.org/abs/1908.07412},
    doi = {10.48550/arXiv.1908.07412},
    urldate = {2024-05-26},
    publisher = {arXiv},
    author = {Qiao, Ning and Indiveri, Giacomo and Bartolozzi, Chiara},
    month = aug,
    year = {2019},
    note = {arXiv:1908.07412 [cs]},
}

@article{pazos_synaptic_2025,
    title = {Synaptic and neural behaviours in a standard silicon transistor},
    volume = {640},
    copyright = {2025 The Author(s)},
    issn = {1476-4687},
    url = {https://www.nature.com/articles/s41586-025-08742-4},
    doi = {10.1038/s41586-025-08742-4},
    language = {en},
    number = {8057},
    urldate = {2026-02-16},
    journal = {Nature},
    publisher = {Nature Publishing Group},
    author = {Pazos, Sebastian and Zhu, Kaichen and Villena, Marco A. and Alharbi, Osamah and Zheng, Wenwen and Shen, Yaqing and Yuan, Yue and Ping, Yue and Lanza, Mario},
    month = apr,
    year = {2025},
    pages = {69--76},
}

@inproceedings{atia_hybrid_2025,
    title = {Hybrid {Integration} of {Rram} and {Cmos} {Circuits} {Using} {Open}-{Source} 130-nm {PDK}},
    issn = {2159-1679},
    url = {https://ieeexplore.ieee.org/abstract/document/11322496},
    doi = {10.1109/ICM66518.2025.11322496},
    urldate = {2026-02-16},
    booktitle = {2025 37th {International} {Conference} on {Microelectronics} ({ICM})},
    author = {Atia, Maryam and Ismail, Mohammed and Alhawari, Mohammad},
    month = dec,
    year = {2025},
    note = {ISSN: 2159-1679},
    pages = {1--6},
}

@inproceedings{gomony_achieving_2024,
    address = {New York, NY, USA},
    series = {{DAC} '24},
    title = {Achieving {PetaOps}/{W} {Edge}-{AI} {Processing}},
    isbn = {979-8-4007-0601-1},
    shorttitle = {Invited},
    url = {https://dl.acm.org/doi/10.1145/3649329.3689623},
    doi = {10.1145/3649329.3689623},
    urldate = {2026-02-23},
    booktitle = {Proceedings of the 61st {ACM}/{IEEE} {Design} {Automation} {Conference}},
    publisher = {Association for Computing Machinery},
    author = {Gomony, Manil Dev and Ahn, Bas and Luiken, Rick and Biyani, Yashvardhan and Gebregiorgis, Anteneh and Laborieux, Axel and Zenke, Friedemann and Hamdioui, Said and Corporaal, Henk},
    month = nov,
    year = {2024},
    pages = {1--4},
}

@book{liu_analog_2002,
    address = {Cambridge, Mass.},
    series = {A {Bradford} book},
    title = {Analog {VLSI}},
    isbn = {978-0-262-12255-9},
    shorttitle = {Analog {VLSI}},
    language = {en},
    publisher = {The MIT Press},
    author = {Liu, Shih-Chii},
    month = nov,
    year = {2002},
}

\end{document}